\documentclass[12pt,preprint]{aastex}
\usepackage{psfig}
\newcommand{\cm}{{\rm\,cm}}
\newcommand{\kms}{{\rm\,km\,s^{-1}}}

\newcommand{\gcm}{{\rm\,g\,cm^{-2}}}

\newcommand{\beq}{\begin{equation}}
\newcommand{\eeq}{\end{equation}}

\newcommand{\bea}{\begin{eqnarray}}
\newcommand{\eea}{\end{eqnarray}}

\newcommand{\beal}{\begin{mathletters}\begin{eqnarray}}
\newcommand{\eeal}{\end{eqnarray}\end{mathletters}}

\newcommand{\nh}{{N_{\rm H}}}
\newcommand{\tho}{\theta_{\rm open}}
\newcommand{\thv}{\theta_{\rm view}}

\slugcomment{Submitted to AJ}
\shortauthors{Belle, Sanghi, Howell, \& Holberg}
\shorttitle{CB Disks}
\begin{document}

\title{A Preliminary Observational Search for Circumbinary Disks
Around Cataclysmic Variables\footnote{ Based on observations made with
the NASA/ESA Hubble Space Telescope, obtained from the Data Archive at
the Space Telescope Science Institute, which is operated by the
Association of Universities for Research in Astronomy, Inc., under
NASA contract NAS 5-26555. This work was started while three of the
authors (KEB, NS, SBH) were at the Planetary Science Institute,
Tucson, AZ.}}

\author{Kunegunda E. Belle\altaffilmark{2}, Neeru Sanghi\altaffilmark{3}, 
Steve B. Howell\altaffilmark{4}, J. B. Holberg\altaffilmark{5},
Peter T. Williams\altaffilmark{2}}
\altaffiltext{2}{X-2 MS P225, Los Alamos National
Laboratory, Los Alamos, NM 87545; belle@lanl.gov, ptw@lanl.gov}
\altaffiltext{3}{Dept. of Astronomy and Dept. of Physics, 
University of Arizona, Tucson, AZ 85721; nsanghi@email.arizona.edu}
\altaffiltext{4}{WIYN Observatory \& NOAO, P.O. Box 26732,
950 N. Cherry Ave., Tucson, AZ 85726-6732; howell@noao.edu}
\altaffiltext{5}{Lunar and Planetary Laboratory, University of Arizona, 
Tucson, AZ 85721; holberg@argus.lpl.arizona.edu}

\begin{abstract}
Circumbinary (CB) disks have been proposed as a mechanism to extract
orbital angular momentum from cataclysmic variables (CVs) during their
evolution.  As proposed by Taam \& Spruit, these disks extend outwards
to several a.u. and should be detected observationally via their
infrared flux or by absorption lines in the ultraviolet spectra of the
CV.  We have made use of archival {\it HST}/STIS spectra as well as
our own near-IR imaging to search for observational evidence of such
CB disks in seven CVs.  Based on the null result, we place an upper
limit on the column density of the disk of $\nh\sim10^{17}\cm^{-2}$.
\end{abstract}

\keywords{novae,cataclysmic variables --- ISM: clouds --- 
stars: individual(QU Carinae, V592 Cassiopeiae, EX Hydrae, DI Lacertae,  
TZ Persei, V3885 Sagittarii, IX Velorum)}

\section{INTRODUCTION}\label{intro}
Cataclysmic variables (CVs) are mass transferring binary star systems
consisting of a white dwarf primary star and red dwarf secondary star.
The evolutionary path from detached binary star system to close,
interacting binary star system is greatly affected by orbital angular
momentum loss (OAML), of which there must be significant amounts in
order to transform the originally long period progenitor into an
interacting binary \citep[see][]{how01}.  A large amount of the OAML
occurs during the common envelope (CE) phase of the CV evolution.

Although the exact mechanism for OAML is still a matter of debate,
several theories have been advanced to explain the observed orbital
period-mass transfer distribution of CVs.  Generally, magnetic braking
is invoked for long period systems ($P_{\rm orb}>3$ hr), and although
these models are marginally successful in explaining the observed
distributions, they predict a strong correlation between mass transfer
rate ($\dot{M}$) and orbital period, along with an excessive number of
ultra-short period, very low luminosity systems; neither the stringent
$\dot{M}-P_{\rm orb}$ correlation, nor the preponderance of low
luminosity systems are observed.  The discord between theory and
observation could be due to a number of different factors, the most
intriguing of which is that magnetic braking may not be solely
responsible for OAML, but rather angular momentum may be extracted
from the binary system via a circumbinary (CB) disk
\citep{spr01,taa01,dub02,taa03}.  The gravitational torque exerted by
the CB disk on the binary will cause significant OAML leading to an
accelerated period evolution and allowing the secondary star to
completely dissolve within a finite time (less than a Hubble time).
These two consequences are the attraction of the CB disk theory, as
they would cause a broad range of $\dot{M}$ for any given period and
remove the need for a large population of faint CVs.

\begin{figure}
\plotone{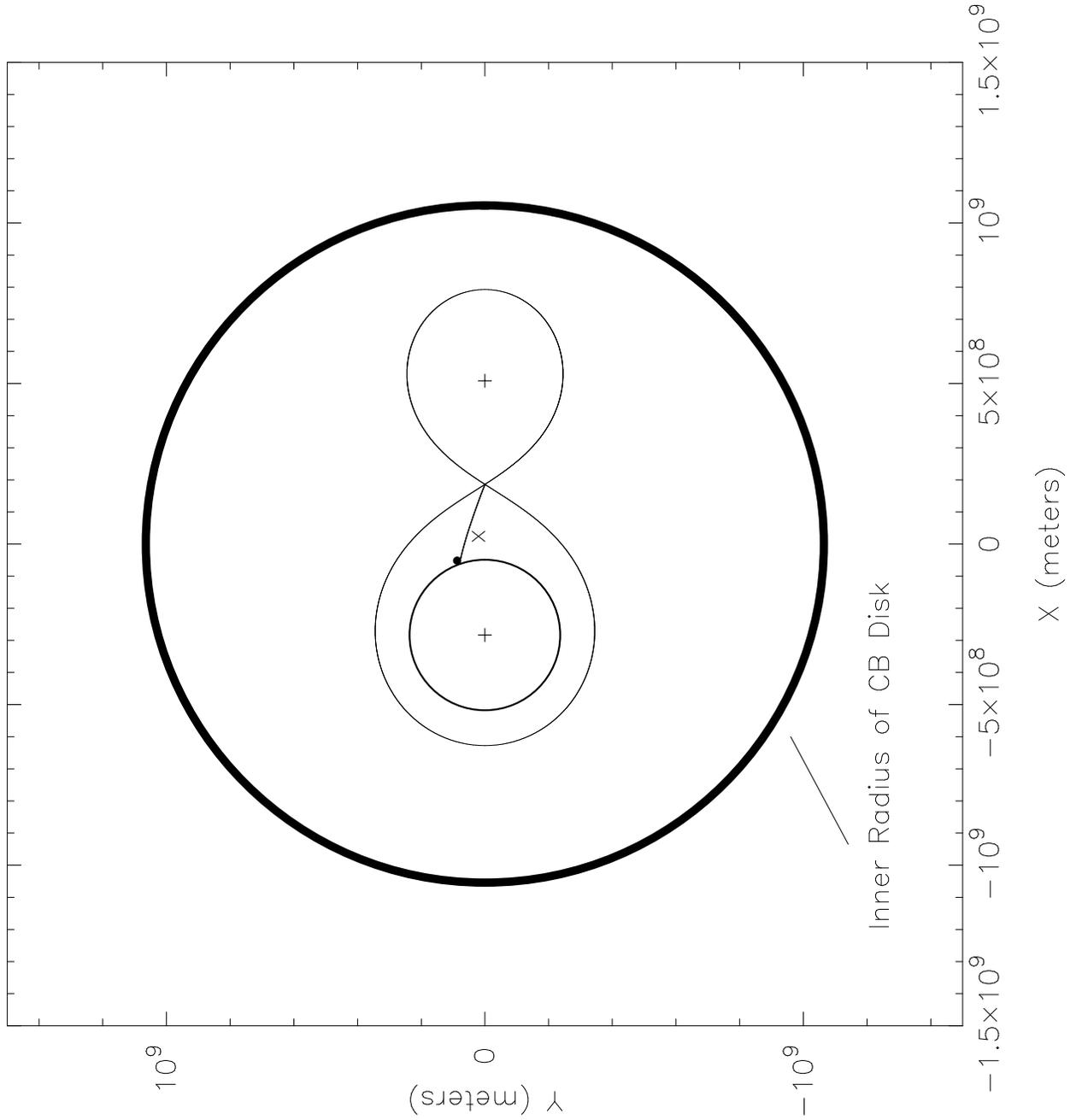}
\caption{A schematic of a CV orbital plane showing the inner edge of a
model CB disk.  The inner disk starts near $10^{11}$ cm, is warm and
optically thick.  The outer edge is cold, optically thin, and extends
out to several A.U. \label{cb1} }
\end{figure}

If CB disks exist, they likely begin formation during the CE phase and
acquire material via mass outflow from the white dwarf (novae) or
accretion disk (winds and outbursts) \citep{spr01}. For the CB disk to
become a significant source of OAML, the mass input rate into the CB
disk must be $10^{-5}$ to 0.01 of the material transferred from the
secondary to the white dwarf throughout the CB disk lifetime.  CB
disks are assumed to be geometrically thin with a pressure scale
height to radius ratio of $H/R\sim0.03$, have a viscosity
$\alpha\la0.01$, and of Keplerian structure.  \citet{dub02} have
proposed that the inner edge is truncated due to tidal forces from the
binary, has a radius near $1\times10^{11}\cm$, effective temperature
$T\sim3000$ K, and is optically thick, while the majority of the outer
disk is cool, $T<1000$ K, optically thin, and can extend out to
several a.u.  Continuum contribution comes from dust emission, with
opacities due to dust grains, and includes contributions from
iron-poor opacities at $T\la1500$ K where dust particles exist.  Based
on models of protoplanetary disks \citep[e.g.,][]{bel97},
\citet{dub02} find that their CB disks reach a maximum vertical height
above the plane of the disk at a radius less than the outer edge of
the disk.  However, the exact physical height of the outer edge of the
CB disk is not known (R. Taam, private communication).
Figure~\ref{cb1} shows a schematic model of a CB disk based on
specifications given in \citet{dub02}.

Two properties of the proposed CB disks will manifest observationally;
the large extent of the CB disk and its surface mass density, $\Sigma
\sim10^2-10^4\, {\rm g}\cm^{-2}$, should make the disk a source of
''ISM-like'' absorption lines appearing at the systemic velocity of
the system, and, if the CB disk emission (from the warm inner edge)
can be considered to be a sum of blackbodies, the flux from the CB
disk will peak near $3\,\micron$ \citep{dub02}.  These properties lend
themselves to detection in two different wavelength regions.  Flux
measurements via near-IR ($1-10\,\micron$) imaging of face-on systems
(offering a full view of the CB disk) may detect the IR flux of the CB
disk.  A second detection technique is to search for the ISM-like
absorption lines produced by the CB disk in UV spectra of the binary
systems.  {\it Hubble Space Telescope}/ Space Telescope Imaging
Spectrograph ({\it HST}/STIS) data are ideal for this as the high
spectral resolution will enable detection and measurement of ISM-like
lines whose radial velocities may then be compared to the space motion
of the binary.  This technique is best done for CVs with inclinations
that match the opening angle of the CB disk, as the binary is then
observed through the plane of the CB disk.  The theoretical work
\citep[e.g.,][]{taa01} predicts that the higher mass transfer rate CVs
are the best candidates for a CB disk, as material may continuously be
entering the CB disk, and may still retain some vertical extent above
the disk (i.e., the CB disk has not had time to ``flatten out'').
Thus, we have begun an observational search for CB disks with long
period nova-likes and old novae.

\section{FEASIBILITY OF CB DISK DETECTION}

While we have chosen CV candidates that should be ideal for detecting
the proposed CB disks, the feasibility of disk detection depends quite
a bit on the unknown CB disk structure.  Here we discuss four
different models for the disk structure - isothermal, isentropic,
constant density scale height, and constant pressure scale height -
and their significance for detecting a CB disk via their absorption
signatures.

\subsection{Key Assumptions}
The CB disk is expected to extend for several orders of
magnitude in barycentric distance, $R$, from the binary \citep{spr01}.
The dependence of the vertically-integrated surface mass density,
$\Sigma$, on radius, $R$, is taken from \citet{taa01}, particularly
Figure~8 in that paper, from which we approximate
\beq
\Sigma = \Sigma_{\rm in} \left({R \over R_{\rm in}}\right)^{-1}.
\label{eq:sigma}
\eeq

Here $\Sigma_{\rm in} = 10^4\gcm$, and $R$ extends from
$R_{\rm in}$ to $R_{\rm out}$, which we take to be $10^{11} \,{\rm
cm}$ and $10^{14}\,{\rm cm}$, respectively \citep{dub02}.  Unlike a
Shakura-Sunyaev disk, the opening angle, $H/R$, is expected to be
roughly constant.  We choose $H/R = \tho \simeq 0.03 \simeq
1.7^\circ$, taken from Figure~2 of \citet{dub02}.

This value for $\tho$, as well as the relation for $\Sigma$ in
eq.~({\ref{eq:sigma}}), are assumed to be the same for each CV
system. The difference between each system is then contained wholly in
the inclination, $i$, or equivalently, the parameter $\thv$, the
viewing angle, where $90^\circ - i = \thv$.  Note that the smallest
viewing angle in our sample occurs with EX Hya, where $\thv=12^\circ \pm
1^\circ$, which corresponds to a height above the midplane of roughly
$z/H \sim 7$, where $z=R \sin \thv \simeq R \thv$ is the height above
the disk.

\subsection{Vertical Disk Profile}
In order to estimate the density above the midplane, we need a model
for the vertical density profile of the disk.  If we are within one
pressure scale height of the midplane, then the uncertainty in the
density is relatively small, but the theoretical uncertainty becomes
much greater for higher $|z|$.  Various processes, such as turbulent
transport, transport by sound waves, and so forth, affect the vertical
entropy profile, which in turn determines the vertical density
profile. In hot disks, there is the further potential complication of
winds being driven off the disk or a hot corona.  Even in relatively
cool disks there may also be photoevaporation off the disk by the
illumination from the central binary.  We leave these considerations
to future theoretical work on the subject of CB disks.

In calculating a vertical density profile for a disk, it is standard
to assume a plane-parallel atmosphere.  Note that in this case there
is a vertical gravity in the disk that increases linearly with $z$:
\beq
g_{\rm eff} = \left( {G M_{\rm binary} \over R^3} \right) z = \Omega_K^2 z,
\label{eq:grav}
\eeq
where $\Omega_K$ is the Keplerian angular velocity,
so that the equation of vertical hydrostatic equilibrium is
\beq
\partial_z P = - \rho \Omega_K^2 z.
\label{eq:hydroeq}
\eeq 
This equation can not be solved without some additional physics to
inform us of the vertical entropy profile, such as the local heating
rate, $q+$, due to turbulent dissipation, and the radiative and
turbulent transport of energy out of the disk.  Some insight into this
problem is provided by theoretical work \citep[e.g.,][]{cas93}, but
uncertainties remain, especially at altitudes of several scale
heights.

First let us comment on the disk thickness (or half-thickness), $H$.
From a disk theorist's perspective, $H$ is usually taken to be a rough
quantity that comes about in basic mixing-length-theory
arguments. Given a location in the disk, $H$ is determined by the
sound speed, which in turn is determined by the temperature. All of
this is tied together by standard thin-disk theory, in which
quantities are taken to be functions of $R$ only, so that the vertical
structure appears only through some rather rough approximations.

It is common to refer to this $H$ as ``the'' pressure scale
height, but when considering the vertical structure in greater detail,
this definition is no longer sufficient.  The local vertical pressure
scale height, $h_P$, defined as $h_P^{-1} = -(\partial_z P)/P$, may
vary significantly with height, $z$, above the midplane; in fact it
becomes formally infinite at the midplane. So, from the operational
standpoint of an observer, there may be some significant ambiguity in
defining $H$, which we need to address.  While there have been
theoretical studies of the vertical density profile for certain disks,
we are not aware of any theoretical treatment of the vertical density
profile for CB disks that extends to several scale heights
above the midplane.  What we offer here is a few simple models for the
disk structure. Hopefully future theoretical and observational work
will further constrain these models.

Mathematically, one option for defining $H$ is that it is the
normalized second moment of the density,
\beq
\Sigma H_1^2 = \int_{-\infty}^{+\infty} \rho z^2 \, dz.
\eeq
An alternative definition of $H$ that is perhaps of more use to the
observer is in terms of the photospheric surface $\tau = 2/3$.  We
also note that \citet{bel97} define $H$ in terms of the ``midplane
pressure scale height'', $c_s / \Omega$.  For the isothermal disk and
the isentropic disk, we take $H = H_1$ as defined above, and for the
constant density scale height and (modified) constant pressure scale
height models, we take $H$ to be equal to the local density and
pressure scale heights, respectively.

\subsection{Model A: Isothermal Disk Atmosphere}
For a disk that is vertically isothermal, at a given radius, $R$, we
have $\rho \sim P$.  The equation of hydrostatic equilibrium,
eq.~(\ref{eq:grav}), then gives $\partial_z \rho \sim - \rho z$,
which has solution $\rho = \rho_0 \exp{(-a z^2)}$.  The local pressure
scale height is $h_P^{-1} = 2 a z$.  If we set $h_P = z = H$,
we find $2aH^2=1$, so that
\beq
\rho = \rho_0 \exp{(-z^2 / 2H^2)}.
\eeq
It is easy to verify that the $H$ in this solution coincides with the
definition for $H_1$ given above, so setting $h_P = z = H$ is
appropriate.  Note that the local pressure scale height $h_P$ (which
in this case is also the local density scale height) is then
\beq
h_P = {H^2 \over |z|},
\eeq
i.e., it diverges at the midplane, is equal to $H$ at $z = \pm H$, and
asymptotes to zero at infinity.

The corresponding surface density, $\Sigma$, for this solution is
\beq
\Sigma = \int_{-\infty}^{+\infty} \rho(z)\, dz = \sqrt{2 \pi} \rho_0 H.
\eeq
In terms of $\Sigma$ and the height above the disk, $z$, the density
becomes
\beq
\rho = {\Sigma \over H \sqrt{2 \pi}} \exp{(-z^2 / 2H^2)}.
\eeq

\subsection{Model B: Isentropic Disk Atmosphere}
For an isentropic disk, we begin with the equation of state $P =
A\rho^\gamma$, where $ A \equiv {P_0 / \rho_0^\gamma}$, and $P_0$ and
$\rho_0$ are fiducial values of the pressure and density, respectively,
taken at the midplane. From equation~({\ref{eq:hydroeq}}) we have that
\beq
{d\rho \over \rho^{2-\gamma}} = -{\Omega_{\rm K}^2 \over \gamma A} z \, dz,
\eeq
which may be integrated to find
\beq
\rho = \rho_0 \left[ 1 - \left({z \over z_{\rm max}}\right)^2 \right]^
{1 \over \gamma - 1}.
\eeq
A very notable aspect of this solution is that $\rho = 0$ for $|z| >
z_{\rm max}$ for some $z_{\rm max}$. Note that $z_{\rm max}$ may be
written as a function of $A$, $\Omega_{\rm K}$, and $\gamma$, although
we do not reproduce the solution here.  Assuming that $\gamma = 5/3$, we
find that
\beq
\Sigma = {3\over 8} \pi \rho_0 z_{\rm max}.
\eeq

We wish to know the $H$ that corresponds to the above solution. Here
the matter is not so simple as for the isothermal case, and we must
take care to define $H$ appropriately, as discussed above. In the case
of $H$ defined in terms of the normalized second moment of the
density, i.e., $H_1$, we find
\beq
H_1 = {z_{\rm max} \over \sqrt{6}} \simeq (2.50)^{-1} z_{\rm max},
\eeq
so that, if we set $H=H_1$, we may now write
\beq
\rho = {\Sigma \over H} {8\over 3 \pi \sqrt{6}} 
\left[ 1 - \left({z \over H\sqrt{6}}\right)^2 \right]^{3/2}
\eeq
for $|z| < H \sqrt{6}$, and $\rho = 0$ otherwise.

\subsection{Model C: Constant Density Scale Height Disk Atmosphere}
The constant density scale height disk is rather trivial. If we define
$H$ as the local density scale height,
\beq
h_\rho^{-1} = - {\partial_z \rho \over \rho} = H,
\eeq
we find that
\beq
\rho = {\Sigma \over 2 H} \exp{(-|z|/H)}.
\eeq

\subsection{Model D: Constant Pressure Scale Height Atmosphere with 
Isothermal Core} The constant pressure scale height atmosphere would
seem to be an attractive model.  However, it presents an unphysical
solution. Namely, if we set
\beq
h_P^{-1} = - {\partial_z P \over P} = H,
\eeq
we find through equation~({\ref{eq:hydroeq}}) that
\beq
\rho \sim {H \over |z|} \exp{(-|z|/H)},
\eeq
so that that density diverges at the midplane, and moreover the vertical
integral of the density, i.e., $\Sigma$, diverges as well.

As an alternative, we consider here an isothermal vertical profile
near the midplane (the ``core''), with a constant pressure scale
height solution ``glued'' onto the isothermal solution at $|z| =
H$. This has the somewhat unsatisfactory result that the vertical
derivative of the density is discontinuous at $|z| = H$, i.e., there
is a ``kink'' in the solution, but this way the divergence mentioned
above is avoided. We then have
\beq
\rho_{|z| < H} = \rho_0 \exp{(-z^2 / 2H^2)}
\eeq
and 
\beq
\rho_{|z| > H} = \rho_0 e^{1/2} {H \over |z|} \exp{(-|z|/H)}
\eeq
where $\rho_0$ is determined by the error function and the exponential
integral, yielding the numerical value
\beq
\rho_0 \simeq (0.4107) {\Sigma \over H}.
\eeq
We may then write
\beq
\rho_{|z| < H} =  (0.4107) {\Sigma \over H} \exp{(-z^2 / 2H^2)}
\eeq
and 
\beq
\rho_{|z| > H} =  (0.6773) {\Sigma \over H} {H \over |z|} \exp{(-|z|/H)}.
\eeq

\subsection{Column Densities for the Models}
It is important to note that, in all cases, the solution
$\rho(\Sigma,H,z)$ may be written in the form
\beq
\rho(\Sigma,H,z) = {\Sigma \over H} f(|z|/H)
\label{eq:fform}
\eeq
for some function $f()$.

The column number density of hydrogen, $\nh$, is
\beq
\nh = \int n_{\rm H} \, d\ell
\eeq
where $d\ell$ is an infinitesimal displacement along the line of sight
to the binary. We assume here that the hydrogen mass fraction of the
disk is near unity, so that $n_{\rm H} \simeq \rho / m_{\rm H}$.

In peforming the integral, the key assumptions are: a constant flare
angle $H/R$ with radius $R$, the expression given above in
eq.~(\ref{eq:sigma}) for the dependence of $\Sigma$ upon radius, and
the functional form in eq.~(\ref{eq:fform}) for the density.
Furthermore, we assume that the density drops immediately to zero when
$R$ is outside the range $(R_{\rm in},R_{\rm out})$.  Note that
$d\ell = \sec {\thv}\, dR \approx dR$. We therefore have
\beq
\nh = {1 \over m_{\rm H}}
\int_{R_{\rm in}}^{R_{\rm out}} \rho(\Sigma,R,z) \, dR
\eeq
The argument of $f()$ that appears in the expression for $\rho$ in
eq.~(\ref{eq:fform}), namely $|z|/H$, is a constant, equal to
$\tan\thv/\tan\tho\approx \thv/\tho$.  The choice of $\Sigma \sim
R^{-1}$ is rather fortuituous from the point of view of performing
this integral. Given that $R_{\rm out} \gg R_{\rm in}$ and using the
small-angle approximation, we find
\beq
\nh = {1 \over m_{\rm H}} {f(\eta) \over \tho} \Sigma_{\rm in}.
\eeq
where $\eta \equiv \thv/\tho$. 

The numerical value of the column density for our four models
may then be written as follows.  For Model A, the isothermal model,
\beq
\nh = (8.03\times10^{27}) 
\left({\Sigma_{\rm in} \over 10^4\gcm}\right)  
\left({\tho \over 1.7^\circ}\right)^{-1}
\exp{\left(-\frac{\eta^2} {2}\right)},
\eeq
For Model B, the isentropic model, we find
\beq
\nh = (6.98\times10^{27})  
\left({\Sigma_{\rm in} \over 10^4\gcm}\right)  
\left({\tho \over 1.7^\circ}\right)^{-1} 
\left(1- \frac {\eta^2}{6}\right)^{3/2}.
\eeq
For Model C, the constant density scale height atmosphere,
\beq
\nh = (1.01\times10^{28})  
\left({\Sigma_{\rm in} \over 10^4\gcm}\right)  
\left({\tho \over 1.7^\circ}\right)^{-1} \exp{(-|\eta|)},
\eeq
and for the constant pressure scale height atmosphere with isothermal core,
\beq
\nh = (8.27\times10^{27})  
\left({\Sigma_{\rm in} \over 10^4\gcm}\right)  
\left({\tho \over 1.7^\circ}\right)^{-1} \exp{\left(-\frac{\eta^2} {2}\right)}
\,\,\,\,
{\rm for} \,\,\,\, |\eta| <1  
\eeq
\beq
\nh = (1.36\times10^{28}) 
\left({\Sigma_{\rm in} \over 10^4\gcm}\right)  
\left({\tho \over 1.7^\circ}\right)^{-1} \left(\frac{1}{|\eta|}\right)\, 
\exp{(-|\eta|)}\,\,\,
\,{\rm for} \,\,\,\, |\eta| >1.
\eeq
Figure \ref{diskmodel} displays these values as functions of
$z/H(=\thv/\tho)$.

\subsection{Discussion of the Solutions}
As may be seen in Figure \ref{diskmodel}, the expected density profile
for small values of $|z|/H$ ($|z| < 2H$) does not depend too much
upon the model chosen. Essentially, $\rho$ is expected to be of the
order of $\Sigma / H$, dropping gradually as $|z|$ increases, and this
is what is seen.

The situation changes quite dramatically, however, when we consider
much larger $|z|$, as can be seen in Figure \ref{diskmodel}. As
discussed above, the density in the adiabatic solution drops to zero
at roughly $|z|/H \simeq 2.50$, for $\gamma = 5/3$ and $H = H_1$. The
density remains rather high in the two constant scale height
solutions, dropping exponentially or close to exponentially depending
upon whether it is the density scale height or the pressure scale
height that is kept constant. The isothermal solution is a sort of
compromise between these extremes, dropping as a Gaussian.  It is
unlikely that a CB disk with a structure described by Model B would be
detectable in any CV system that does not have $i\sim90^{\circ}$.  CB
disks with structures described by models A, C, and D, however, would
be detectable in systems with lower inclinations.

\section{OBSERVATIONS}\label{obs}
As this project is an initial investigation for CB disks around CVs, we
chose to look at a concise sample of objects that met our criteria:
bright, high mass transfer systems (including the old nova, DI~Lac);
known inclinations and systemic velocities; and existing 
{\it HST}/STIS spectra that would be retrievable via the Multimission
Archive at Space Telescope (MAST).  The six objects we chose for
spectroscopic investigation are listed in Table~\ref{objects} and the
data sets utilized are given in Table~\ref{log}.  Each {\it HST}/STIS
observation employed the E140M grating in Echelle mode, which spans
the wavelength range $1150-1735$\AA, and has a resolving power of
$R=45,800$, or $\sim0.02$ \AA\ pixel$^{-1}$ throughout each spectrum.
The data were retrieved from the MAST archive and processed using
IRAF/STSDAS software.  We used IRAF/SPLOT to analyze the spectra.

Four of the spectra are of excellent quality: IX~Vel and V3885~Sgr
both have a signal-to-noise ratio of $\sim30$ in the continuum, and 
QU~Car and EX~Hya have $S/N\sim20$ in the continua.  The spectra of 
DI~Lac and TZ~Per are slighty noisier, with signal-to-noise ratios of 5
and 10, respectively.  The DI~Lac spectrum is too noisy shortward of
$1200$\AA\ to properly identify any spectral features below this
wavelength.

We also present near-IR $JHKL'$ band photometric data of the nova-like
CV, V592~Cassiopeia.  V592~Cas was observed in $JHK$ at the Wyoming
Infrared Observatory (WIRO) in 1997 \citep{cia98} and in $L'$ at the
Fred Lawrence Whipple Observatory 1.2 m telescope on Mt. Hopkins for a
total of 5 minutes on 1999 January 1 UT.  The $JHKL'$ data were obtained
as part of a campaign to study IR properties of CV secondaries.
Additional $JHK$ observations of V592~Cas have been secured at the
United Kingdom IR Telescope on Mauna Kea and are in agreement with the
WIRO data.

\section{ANALYSIS AND DISCUSSION}\label{disc}
For each spectrum, we identified absorption lines of ionic species
commonly associated with the ISM and CV systems, such as C, N, Mg, Si,
and S.  We considered absorption lines that are above the $2\sigma$
error level in flux and that have full width at half maximum (FWHM)
values $<0.2$ \AA, as absorption lines with larger FWHM values would
not be associated with the slowly rotating CB disk or the ISM.
Absorption lines were fit with a single Gaussian function to determine
line centers and equivalent widths (EWs). Some absorption lines are
composed of well-separated multiple absorption components (for
example, the absorption lines of QU~Car); for these we measured the
line centers of the components individually.  At the resolution of the
E140M grating, our spectra are not sufficient to resolve closely
spaced ISM features; for absorption lines containing unresolvable
components, we measured the lines with a single Gaussian.  Radial
velocities for all measured absorption components were then calculated
and are given in Table~\ref{all_tab}, along with the EWs of the
absorption lines.  The measurement errors on the radial velocities for
narrow lines that are unblended and unsaturated are within the
velocity resolution of the spectrum, which is typically $\sim3\kms$.
Lines with EW $\,>100$ m\AA\ are likely saturated or close blends of
two strong components.  In such circumstances, it is difficult to
assign a meaningful centroid velocity to these features; we therefore
assign uncertainties of $\pm7\kms$ to the radial velocities calculated
for these lines.  Absorption lines of excited state ions are noted in
the table by an asterisk.

\subsection{ISM Lines of Sight}
Many of the identified absorption lines will be due to absorption by
the ISM, in particular the local interstellar cloud (LIC); the warm,
partially ionized cloud in which the Sun resides.  The heliocentric
velocity of the LIC is $25.7\kms$ towards the Galactic coordinates
$l=186\fdg1$, $b=-16\fdg4$ \citep[e.g.,][]{woo02}.  Evidence of the
LIC is frequently observed over a large fraction of the sky as a
distinct range of velocities associated with ISM absorption components
present in high dispersion UV spectra of nearby stars
\citep[e.g.,][]{lal92,hol98,woo02,kim03}.  In addition to the LIC,
many other velocity components are also observed in certain
directions.  The LIC and these other components are commonly discussed
in terms of an assemblage of warm interstellar clouds within 30 pc of
the sun.  A good up-to-date and lucid description of these ISM
velocity components is provided in \citet[hereafter FGW]{fri02}.  In
Table~\ref{objects}, we give the velocity of the LIC in the direction
of each of our candidate objects.

The LIC velocity is easily recognized in the histograms we have
created for the ISM-like absorption lines found in the spectra of the
six CV systems we studied.  These histograms are shown in
Figures~\ref{dilac_hist} - \ref{v3885_hist}.  For four of the six
objects, DI~Lac, IX~Vel, TZ~Per, and QU~Car, the LIC velocity is at or
near the peak velocity in the histograms.  EX~Hya and V3885~Sgr do not
follow suit; the peak in the EX~Hya histogram is $\sim7\kms$ from the
LIC velocity, and neither of the peaks in the V3885~Sgr histogram are
aligned with the LIC velocity.

The line of sight to DI Lac lies in a direction expected to show LIC
components.  However, the projected LIC velocity very nearly coincides
with the systemic velocity of the system, making it difficult to
observationally distinguish either (see Figure~\ref{dilac_hist}).
Complicating matters is the presence of lines due to interstellar
\ion{C}{1}, an ion not present in the LIC because it is easily
destroyed by the ambient stellar UV radiation field.  The presence of
\ion{C}{1} lines indicates that the line of sight to DI Lac passes
through a diffuse cloud with sufficient density to preserve the
\ion{C}{1} ion.  The presence of an unusually dense component of the
ISM is also indicated by the very strong and saturated lines of
\ion{S}{2}.  The mean velocity of the \ion{C}{1} lines and the
\ion{S}{2} lines ($-1.78\kms$) contributes significantly to the peak
in the velocity distribution adjacent to the predicted LIC velocity.
It would be tempting to associate these lines with a CB disk, however,
the velocities are not consistent and the physical conditions in the
CB disk are not appropriate to the existence of the \ion{C}{1} ion.
Rather, it appears that the line of sight to DI Lac passes through a
relatively dense interstellar cloud, unrelated to the CV system.
There is no evidence in the spectrum of DI Lac of lines arising from a
CB disk.

The line of sight to EX Hya lies in a direction that shows no LIC
components.  The peak of the radial velocity histogram (Figure
\ref{exhya_hist}) does, however, coincide with the expected velocity
of the Filament structure described in FGW, consistent with several
stars that lie in the same general direction as EX Hya.  There is no
evidence for a CB disk and the line of sight to this star is
consistent with the ISM only.

For IX Vel and QU Car, the line of sight to each object lies in a
direction expected to show LIC components and the peaks of the
histograms coincide with the expected LIC velocity (see Figures
\ref{ixvel_hist} and \ref{qucar_hist}).  The line of sight to IX Vel
is consistent with the ISM only.  The spectrum of QU Car shows
several relatively weak \ion{C}{1} lines, indicating the line of sight
passes through a relatively dense interstellar cloud.  There is no
indication of CB disk related absorption for QU Car.

The line of sight to TZ Per lies in a direction expected to show LIC
components, however, the histogram of velocities (see Figure
\ref{tzper_hist}) shows a broad distribution of velocities.  Like the
spectrum of DI Lac, the spectrum of TZ Per contains \ion{C}{1} lines
as well as relatively strong \ion{S}{2} lines, consequently a similar
interpretation - a line of sight through a dense interstellar cloud -
holds for TZ Per.  Many of the features in the TZ Per spectrum are
either saturated or quite strong, indicating that several unresolved
velocity components are present and contributing to the broad
distribution of velocities.  There is one \ion{C}{1} line appearing
near the systemic velocity, however, this is not sufficient evidence
to support absorption from a CB disk and as discussed above, the
existence of this ion in a CB disk is not probable.

The line of sight to V3885 Sgr contains two distinct, well separated
velocity components (see Figure \ref{v3885_hist}).  The component at
$+2.5\kms$ can be confidently identified with the South Polar Cloud
(see FGW).  The second peak near $20\kms$ corresponds to no previously
described interstellar cloud in the local ISM.  There is no clear
contribution from the LIC or a CB disk.  The line of sight to this
star is consistent with the ISM only.

\subsection{Absorption Line Velocities}
Figures~\ref{di_vel} - \ref{v3_vel} display line velocity profiles for
select absorption lines from the spectra of each of the six CVs.  On
each plot, the vertical dashed line represents the LIC velocity
towards the object, and the vertical dotted line represents the
systemic velocity of the system.  For systems that have systemic
velocities with large errors, the dotted line represents the measured
value.  We selected a sampling of absorption lines to illustrate the
line structure and velocity behavior for each system.  Shown are the
common absorption lines of the \ion{N}{1} $\lambda\lambda
1199.55,1200.22,1200.71$ triplet, lines from several \ion{C}{1}
multiplets, the \ion{S}{2} $\lambda\lambda 1250.58,1253.81,1259.52$
triplet, lines from several \ion{Si}{2} multiplets, along with the
\ion{Fe}{2} $\lambda1608.46$ and \ion{Al}{2} $\lambda 1670.79$
absorption lines (excited transitions are noted with an asterisk).

Figure~\ref{di_vel} displays absorption lines found in the spectrum of
DI~Lac.  Each line exhibits a single velocity structure, however,
because the LIC velocity and systemic velocity are separated by less
than the velocity resolution of the spectrum, it is not possible to
identify the absorption lines with either velocity.  As discussed in
the previous section and based on the results from the remaining
objects of our study (see below) it is likely that the absorption
lines in the DI~Lac spectrum are ISM in nature.  The 1910 nova
eruption of DI~Lac implies the presence of nova shells, which may be
observed via absorption lines, however, these lines should be
blue-shifted with respect to the systemic velocity.

The absorption lines found in the EX~Hya spectrum (shown in
Figure~\ref{ex_vel}) have a single velocity structure.  They are
shifted $\sim7-10\kms$ with respect to the LIC velocity of
$-14.8\kms$, but are aligned with the Filament structure discussed in
FGW.  The line velocities of IX~Vel (Figure~\ref{ix_vel}) and QU~Car
(Figure~\ref{qu_vel}) are aligned with the LIC velocity.  In each
case, the systemic velocity of the system is clearly not associated
with the absorption line velocities.

TZ~Per (Figure~\ref{tz_vel}) is another system for which the LIC and
systemic velocities fall within the velocity width of the absorption
line profiles.  The two velocities are separated by more than the
velocity resolution of the spectrum, therefore the absorption line
velocities may be identified with the LIC velocity, as it is more
closely aligned with the absorption line centers (the LIC velocity
alignment is also seen in Figure~\ref{tzper_hist}).

The absorption lines in the V3885~Sgr spectrum (Figure~\ref{v3_vel})
have two velocity components; one that is aligned with the South Polar
Cloud (FGW) and one that does not match any known ISM velocity system.
Sagittarius is in the direction of the galactic center, so it is
likely that different clouds along this line of sight are causing
absorption along the line of sight to V3885~Sgr.

\subsection{Infrared Observations}
Although the outer regions of the proposed CB disks are cool and
optically thin and not a source of IR blackbody flux, the warm,
optically thick inner regions of the disk will make a contribution to
the IR flux of a CV system.  Given that the inner region of the CB
disk is predicted to have a temperature of up to $\sim3000$ K and is
optically thick, we approximated a mean temperature of the CB disk
that agrees well with Figure 2 of \citet{dub02} to create a blackbody
model representative of the CB disk.  Determining this temperature is
important, as the peak of the blackbody curve of the CB disk should
cause an observable 'bump' in the spectral energy distribution (SED)
of the CV.

V592~Cas has an inclination of $i=27\degr$, and is therefore ideal for
detecting the full CB disk via IR observations.  In Figure~\ref{fred},
we plot $JHKL'$ points for V592~Cas, along with blackbody curves
representing the secondary star for the system and a steady state
accretion disk.  Also plotted is the sum of the IR flux from all
components of the CV system, including the flux contribution expected
from a CB disk; it is a sum of blackbodies and peaks near
$3\,\micron$.  This model summed spectrum overestimates the IR flux
observed for V592~Cas; the data points do not fall along the model
curve that includes the flux contribution of the CB disk, but rather
follow the blackbody curve of the steady state accretion disk.  It is
apparent that flux from a CB disk is not evident in the IR data of
V592~Cas.

\subsection{Observations of the ISM}
Most of the spectral features observed in the six CV systems studied
can be reasonably assigned to the LIC or other recognized velocity
structures in the local ISM near the Sun.  However, in several CVs (DI
Lac, QU Car, and TZ Per), we do observe unusually strong lines due to
\ion{S}{2} and features due to \ion{C}{1}, indicating that the line of
sight to these stars passes through a dense, warm ISM component where
\ion{C}{1} is shielded from ionizing radiation.  It is regarded as
unlikely that these regions are associated with either CB disks or
other gas related to the CV, as the observed velocities of these
components are distinctly different from the CV system velocity.

\subsection{Evidence for CB Disks?}
In the spectroscopic and photometric data presented here, we have
found no convincing evidence to support the theory of CB disks around
CVs.  The ISM-like lines appearing in the spectra of the six CVs have
velocities consistent with the ISM in the direction of the object.
The two cases in which the systemic velocity fell within the velocity
width of the measured absorption lines (DI~Lac and TZ~Per) were
complicated by the fact that the systemic velocity was close to the
LIC velocity and that the identified lines were unlikely to occur in a
CB disk.  Based on the results of the other four objects, the match of
the systemic velocity is due purely to its proximity to the LIC
velocity.  Evidence for a CB disk was not found in the $JHKL'$
photometric data of V592~Cas.  There is no excess of flux around the
expected peak wavelength of $3~\micron$, and models of a CB disk
overestimate the flux measured in V592~Cas.

In the absence of any absorption features clearly associated with CB
disks, the upper limits for the disk material may be derived based on
3$\sigma$ upper limits for the EWs of \ion{N}{1}, \ion{Si}{2},
\ion{O}{1}, and \ion{S}{2}, evaluated at the systemic velocity in the
spectrum of each CV (except for DI Lac, which exhibits some lines near
the systemic velocity).  The 3$\sigma$ EW values for all lines
measured are very low - they range between 5 m\AA\ - 25 m\AA\ - so we
can use the linear portion of the curve of growth to determine the
species column densities: $N_i =
1.129\times10^{20}(\frac{W_i}{\lambda_i}) (\frac{1}{f_i \lambda_i})$.
Wavelengths and $f$ values are taken from \citet{mor91}.  To calculate
equivalent H column densities, we use typical abundances of each
species observed for the warm phase of the ISM, $A_i$, such that
$\nh = \frac{N_i}{A_i}$.  The best upper limits we calculate are
from the \ion{N}{1} and \ion{O}{1} lines.  For EX Hya we determine a
value of $\nh \sim 3\times10^{17}\cm^{-2}$, for IX Vel $N_{\rm
H} \sim 1\times10^{17}\cm^{-2}$, for QU Car $\nh \sim
2\times10^{17}\cm^{-2}$, for TZ Per $\nh \sim
2\times10^{17}\cm^{-2}$, and for V3885 Sgr $\nh \sim
2\times10^{17}\cm^{-2}$.  

\begin{figure}
\plotone{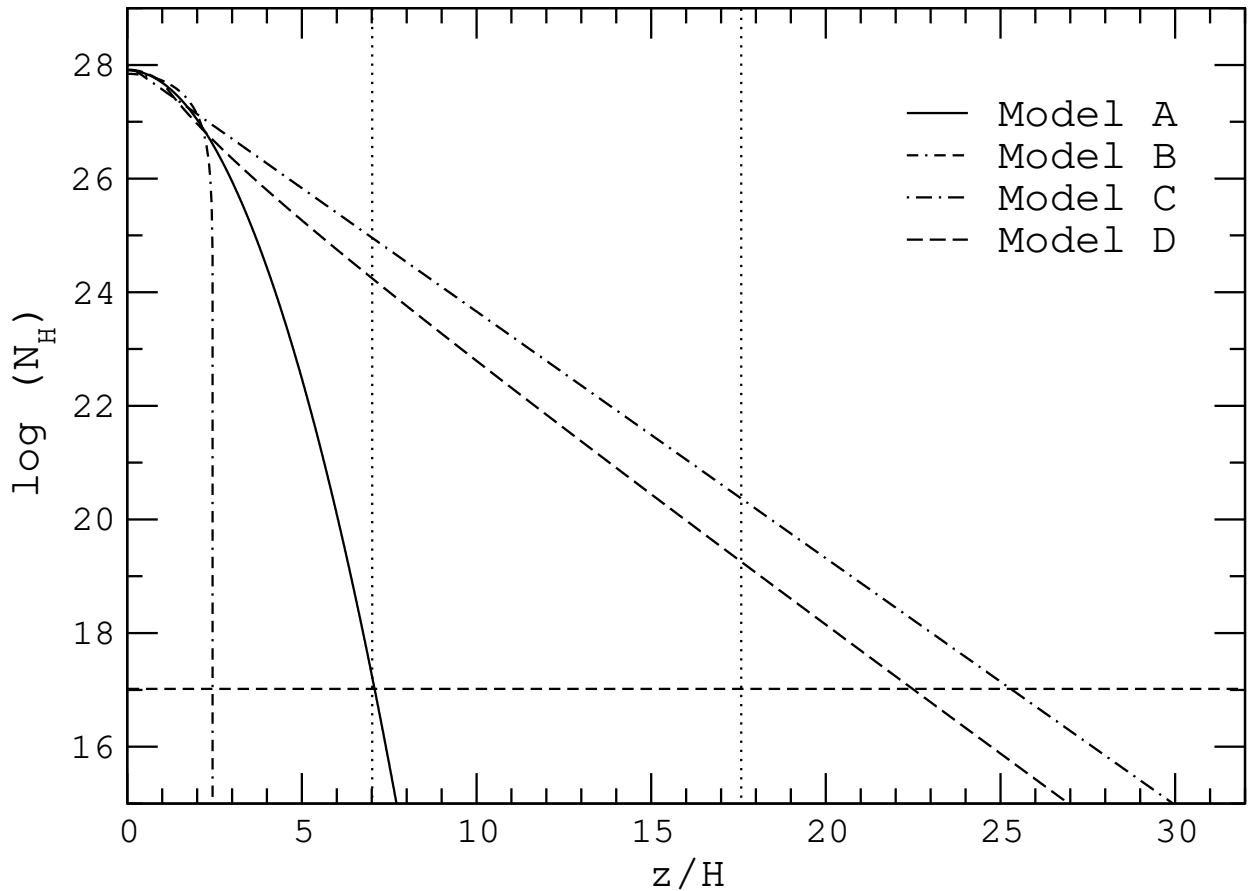}
\caption{Vertical density profiles above the CB disk midplane for disk
structure models A, B, C, and D.  The vertical dotted line at $z/H=7$
represents the value for EX Hya, while the line at $z/H=17$ represents
the z/H value for IX Vel, QU Car, and V3885 Sgr.  The horizontal
dashed line at $\log \nh=17$ represents the $3\sigma$ upper limit for
the column densities, based on measurements of the HST spectra
(discussed in a later section). \label{diskmodel}}
\end{figure}

\begin{figure}
\plotone{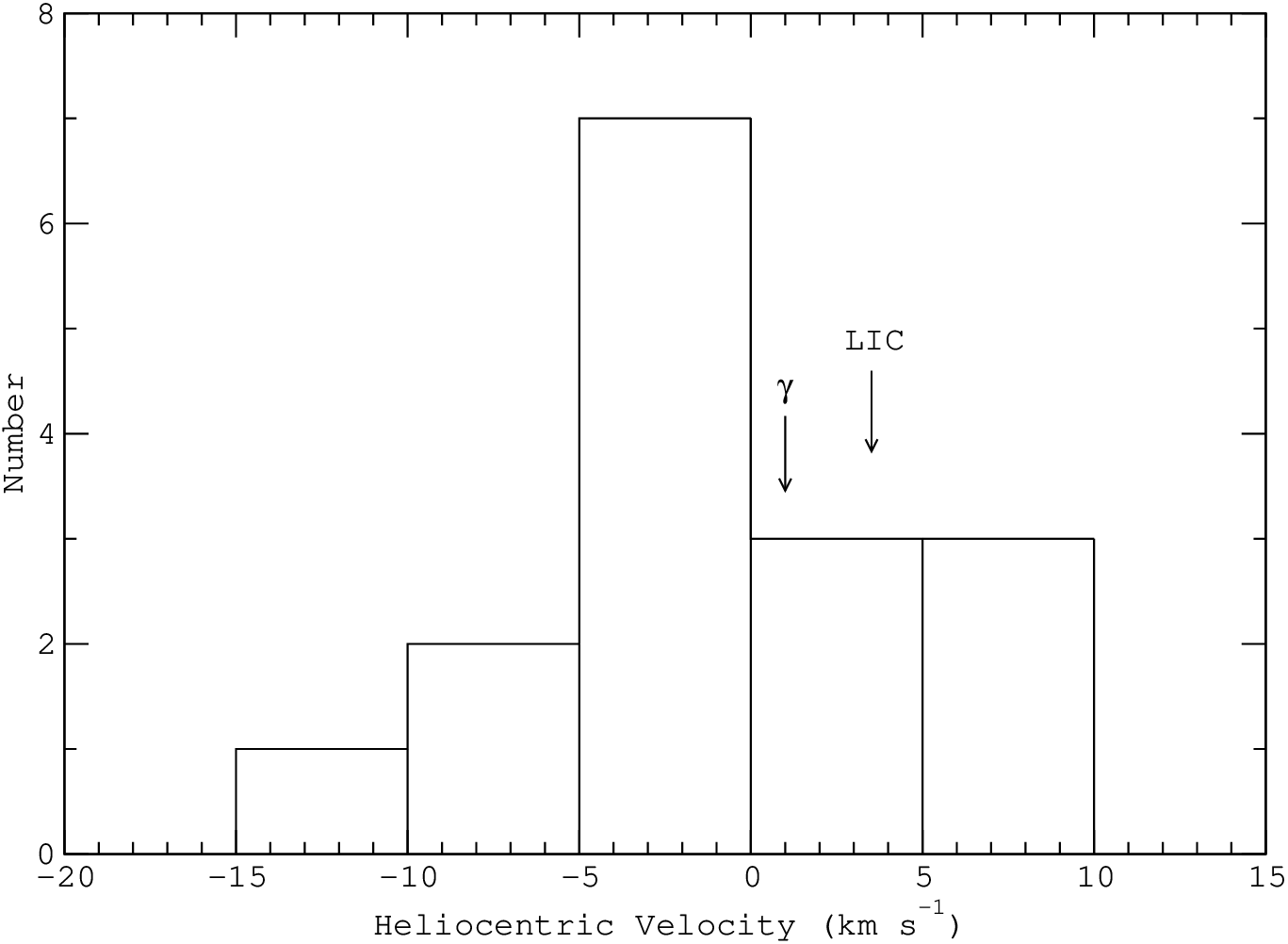}
\caption{Histogram of the radial velocities of ISM-like absorption
lines measured in the spectrum of DI~Lac.  Labeled are the systemic
($\gamma$) and calculated LIC velocities, neither of which match the
peak velocity in the histogram.  \label{dilac_hist}}
\end{figure}

\begin{figure}
\plotone{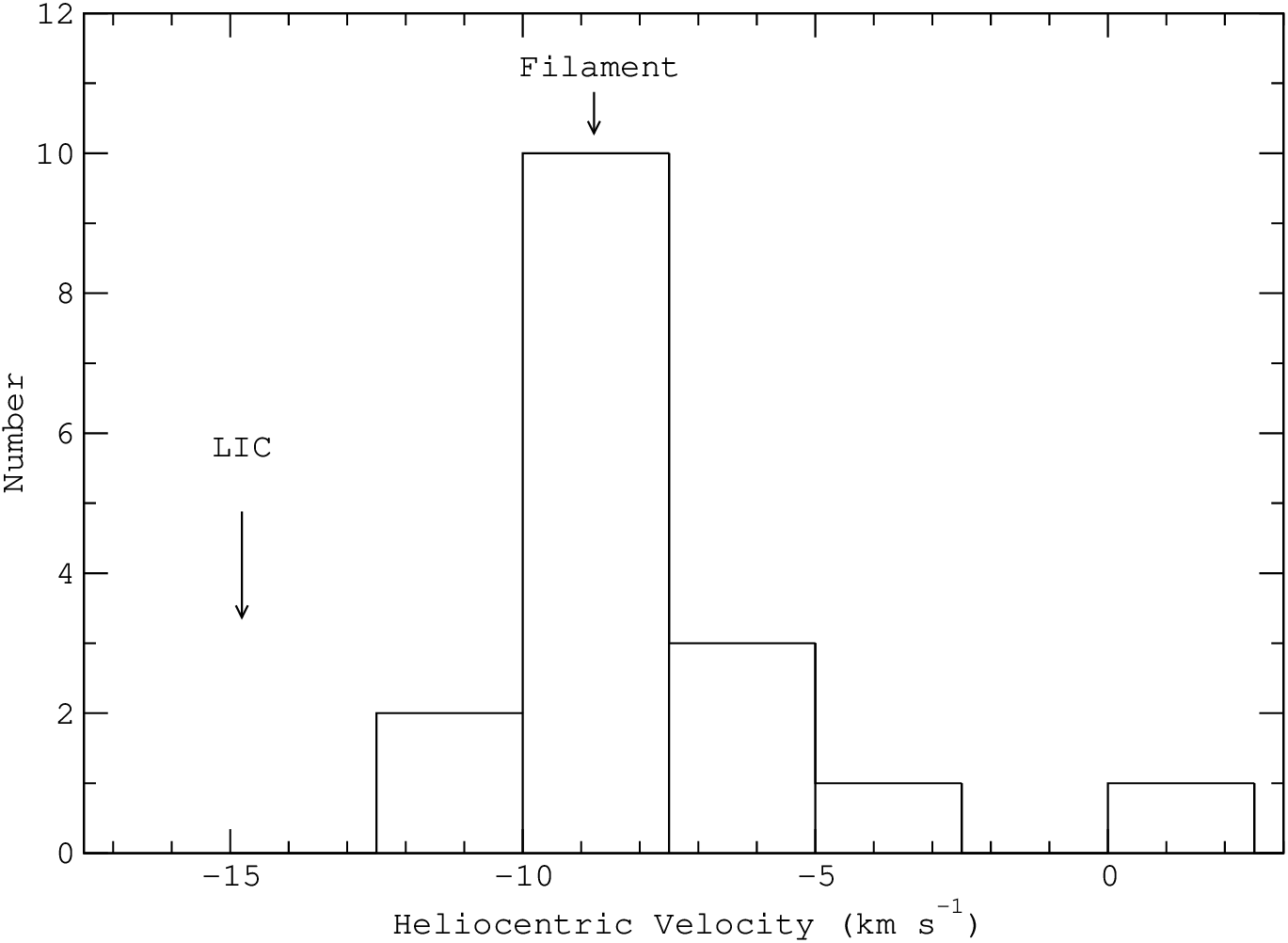}
\caption{Histogram of the radial velocities of ISM-like absorption
lines found in the spectrum of EX~Hya. The LIC velocity is labeled at
$-15\kms$, while the velocity of the Filament structure discussed in
FGW at $-9\kms$ is also labeled. \label{exhya_hist}}
\end{figure}

\begin{figure}
\plotone{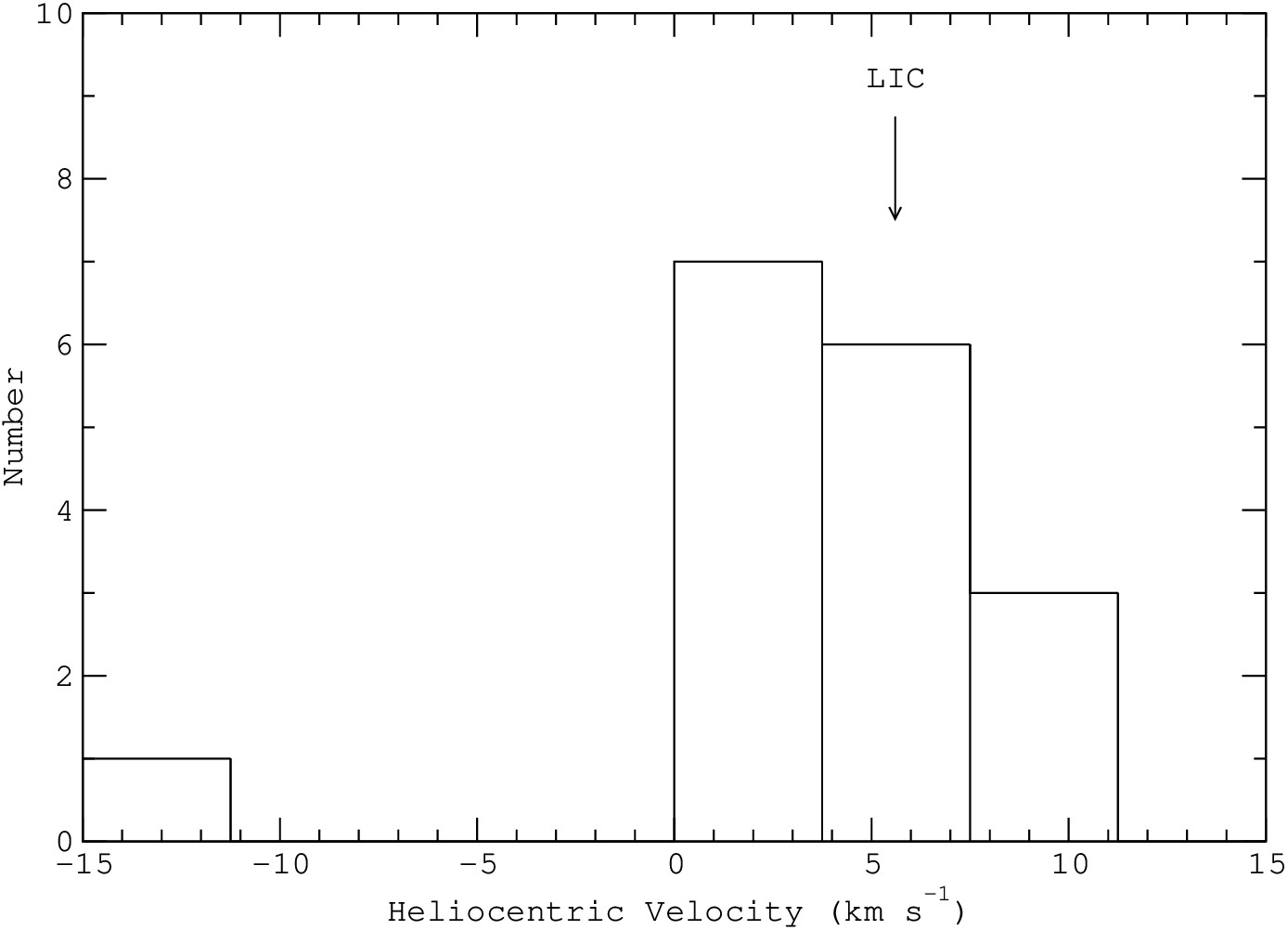}
\caption{Histogram of the radial velocities of ISM-like absorption
lines found in the spectrum of IX~Vel.  The peak of the histogram
matches well with the LIC velocity of $5.6\kms$.    \label{ixvel_hist}}
\end{figure}

\begin{figure}
\plotone{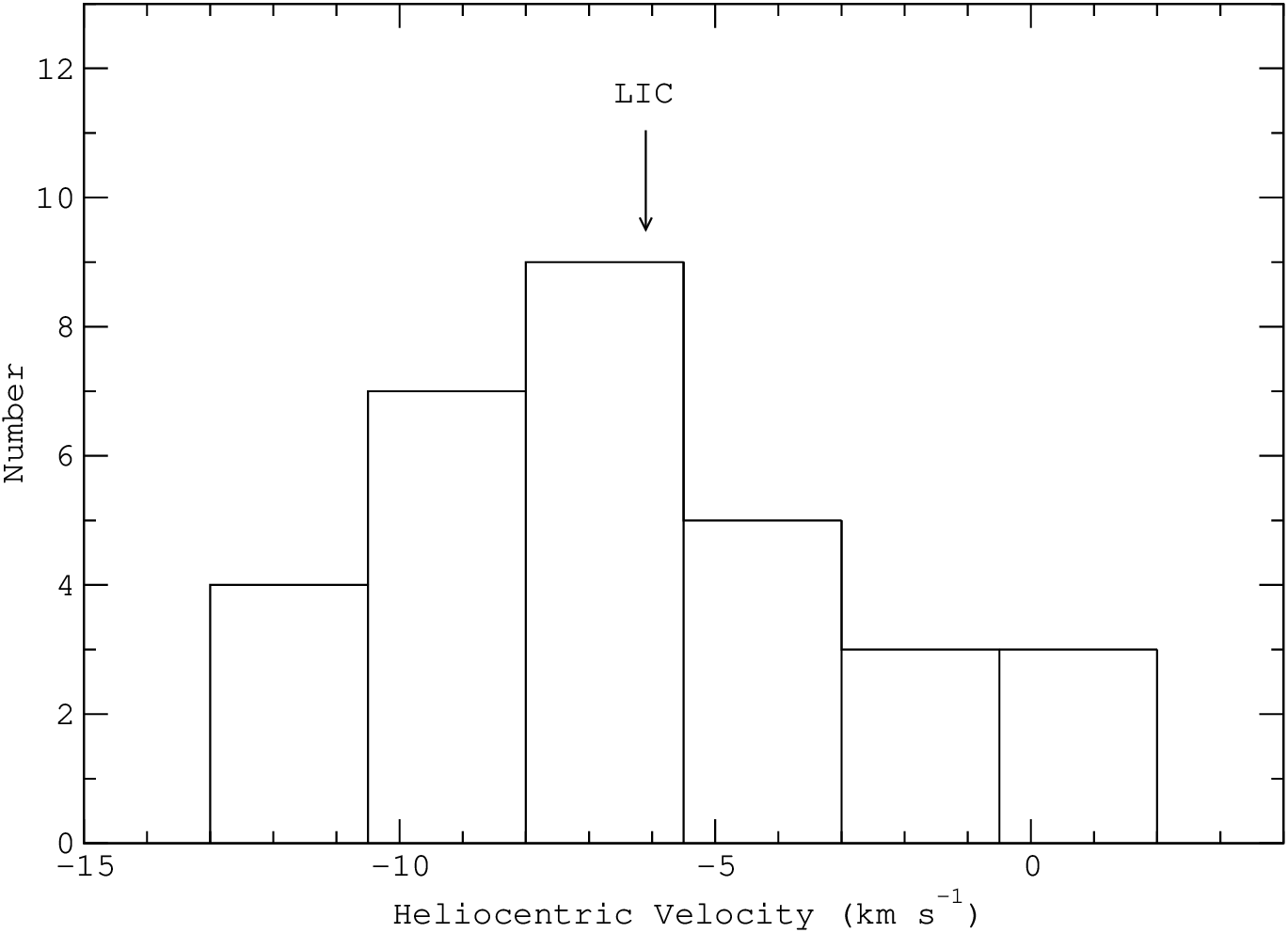}
\caption{Histogram of the radial velocities of ISM-like absorption
lines found in the spectrum of QU~Car.  The peak velocity of the
histogram corresponds to the LIC velocity of $-6.1\kms$.
\label{qucar_hist}}
\end{figure}

\begin{figure}
\plotone{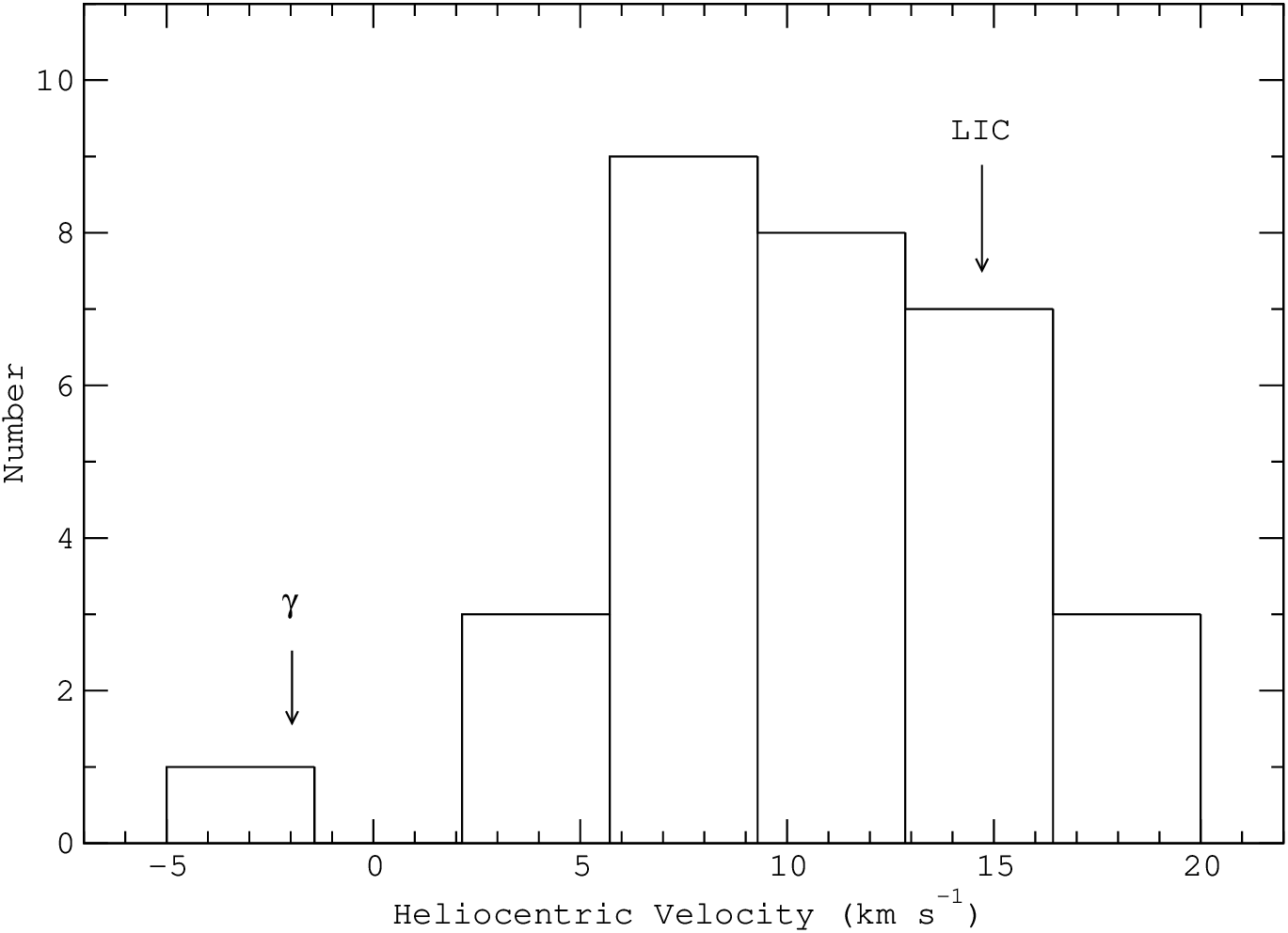}
\caption{Histogram of the radial velocities of ISM-like absorption
lines found in the spectrum of TZ~Per.  There is a grouping of lines
with velocities around $5-15\kms$ - the LIC velocity falls at the high
end of this range.  One absorption line has a velocity that matches
the systemic velocity (labeled as $\gamma$) at $-2\kms$.  \label{tzper_hist}}
\end{figure}

\begin{figure}
\plotone{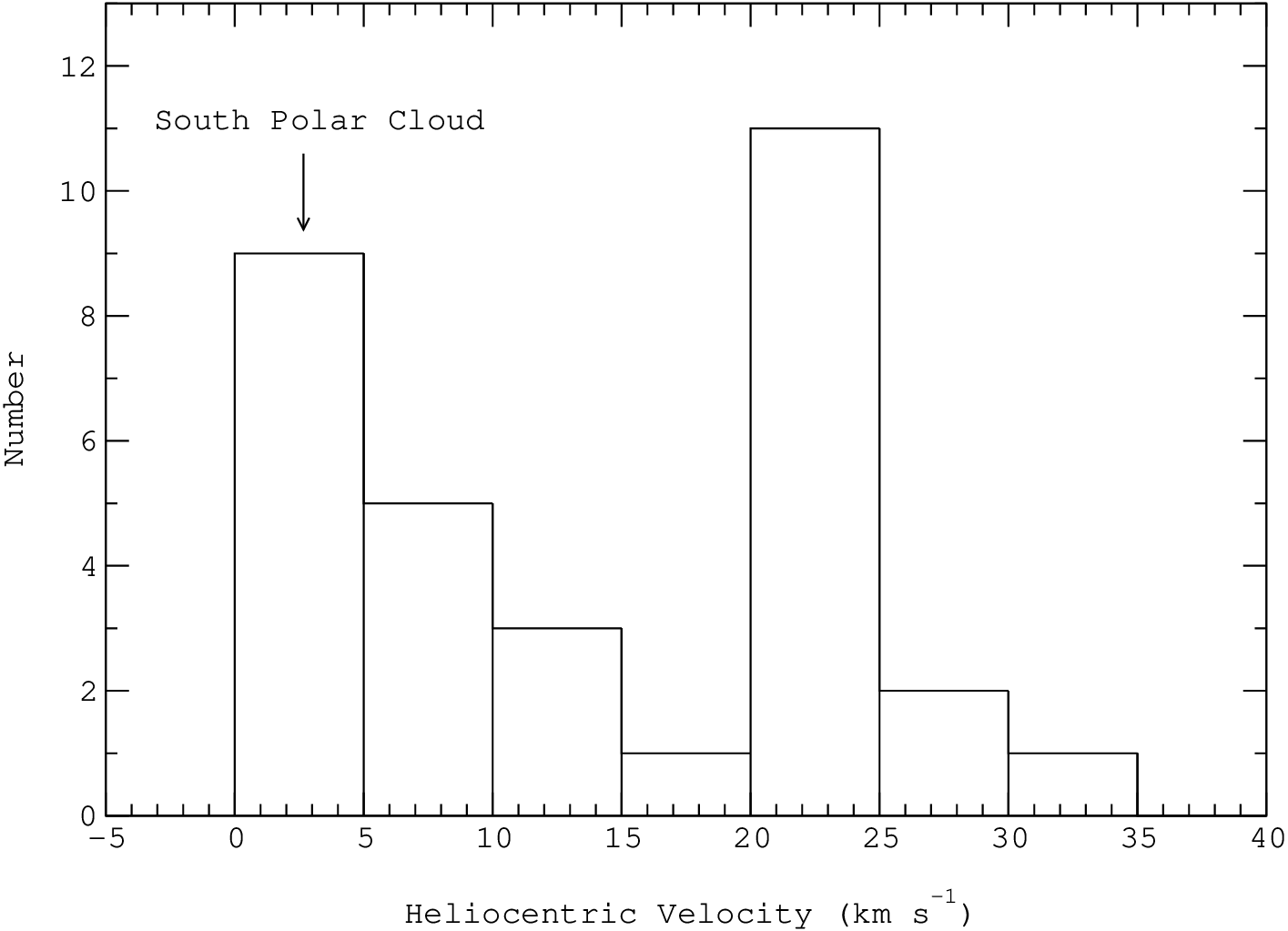}
\caption{Histogram of the radial velocities of ISM-like absorption
lines found in the spectrum of V3885~Sgr.  The two peaks in the
histogram, at $\sim4\kms$ and $\sim22\kms$, represent the two velocity
components of the absorption lines.  The peak at $\sim4\kms$ matches
well with the South Polar Cloud. \label{v3885_hist}}
\end{figure}

\begin{figure}
\plotone{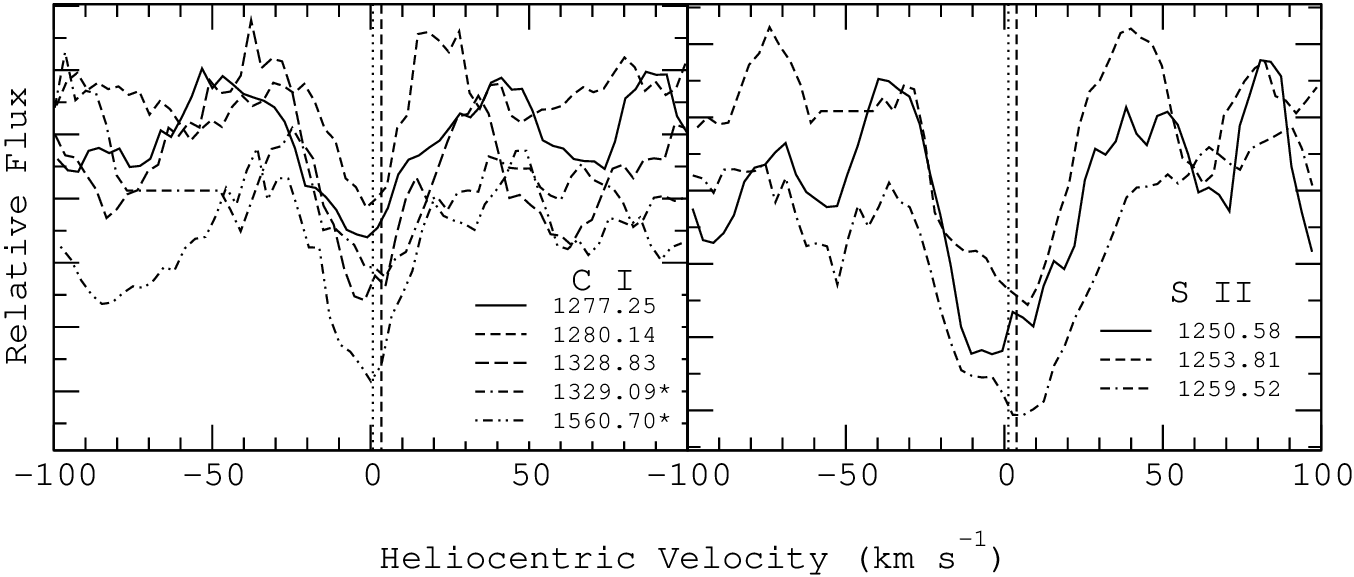}
\caption{Line velocity profiles for absorption lines from four
\ion{C}{1} multiplets and the \ion{S}{2} triplet observed in the DI~Lac 
spectrum ($S/N\sim5$).  The dashed line at $3.5\kms$ represents
the LIC velocity, while the dotted line at $1\kms$ represents the
systemic velocity. \label{di_vel}}
\end{figure}

\begin{figure}
\plotone{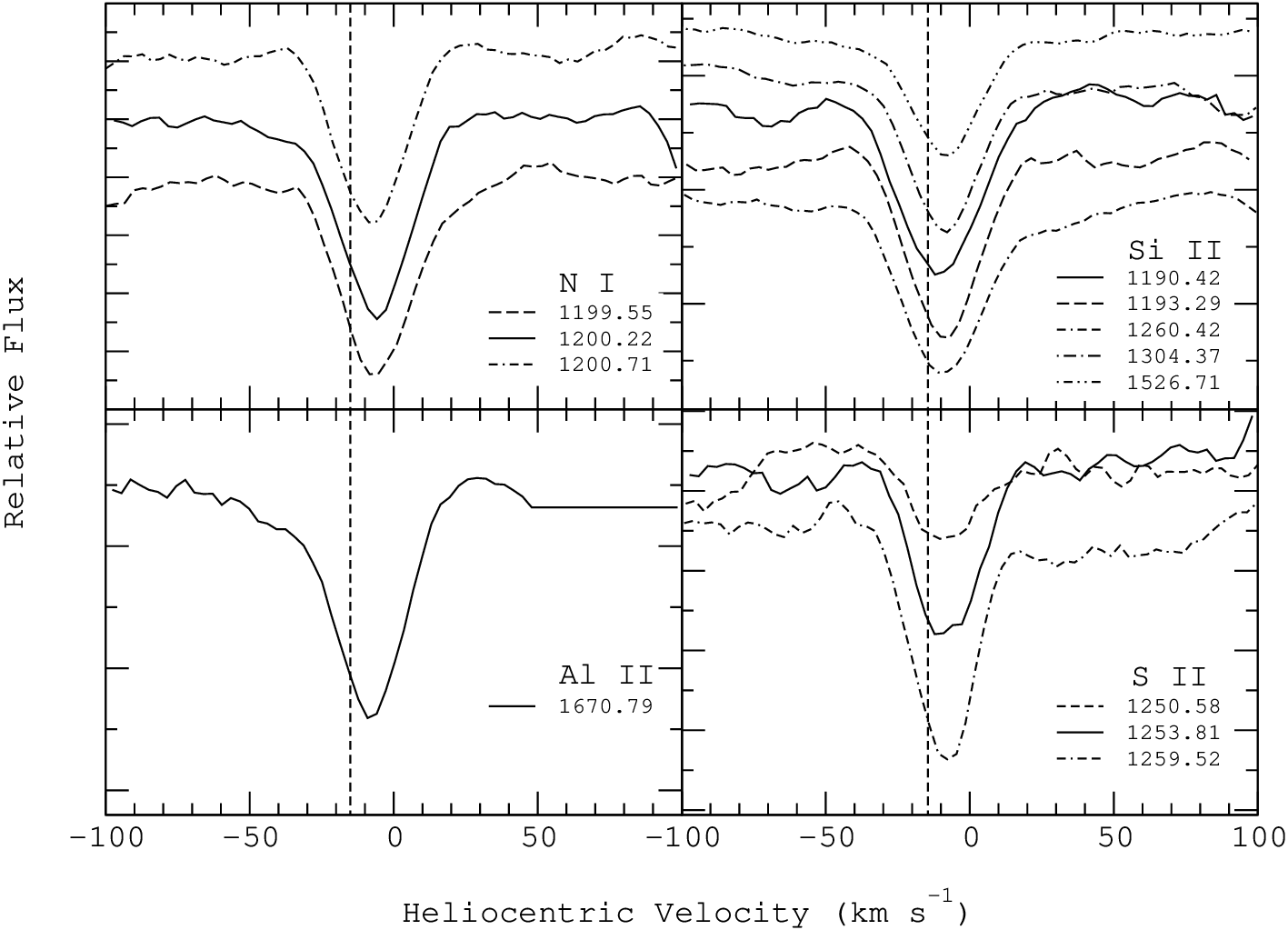}
\caption{Line velocity profiles for the \ion{N}{1} triplet, lines
from several \ion{Si}{2} multiplets, the \ion{Al}{2}$\lambda1670$
line, and the \ion{S}{2} triplet found in the EX~Hya spectrum
($S/N\sim20$).  The dashed line at $-14.8\kms$ represents the LIC
velocity.  \label{ex_vel}}
\end{figure}

\begin{figure}
\plotone{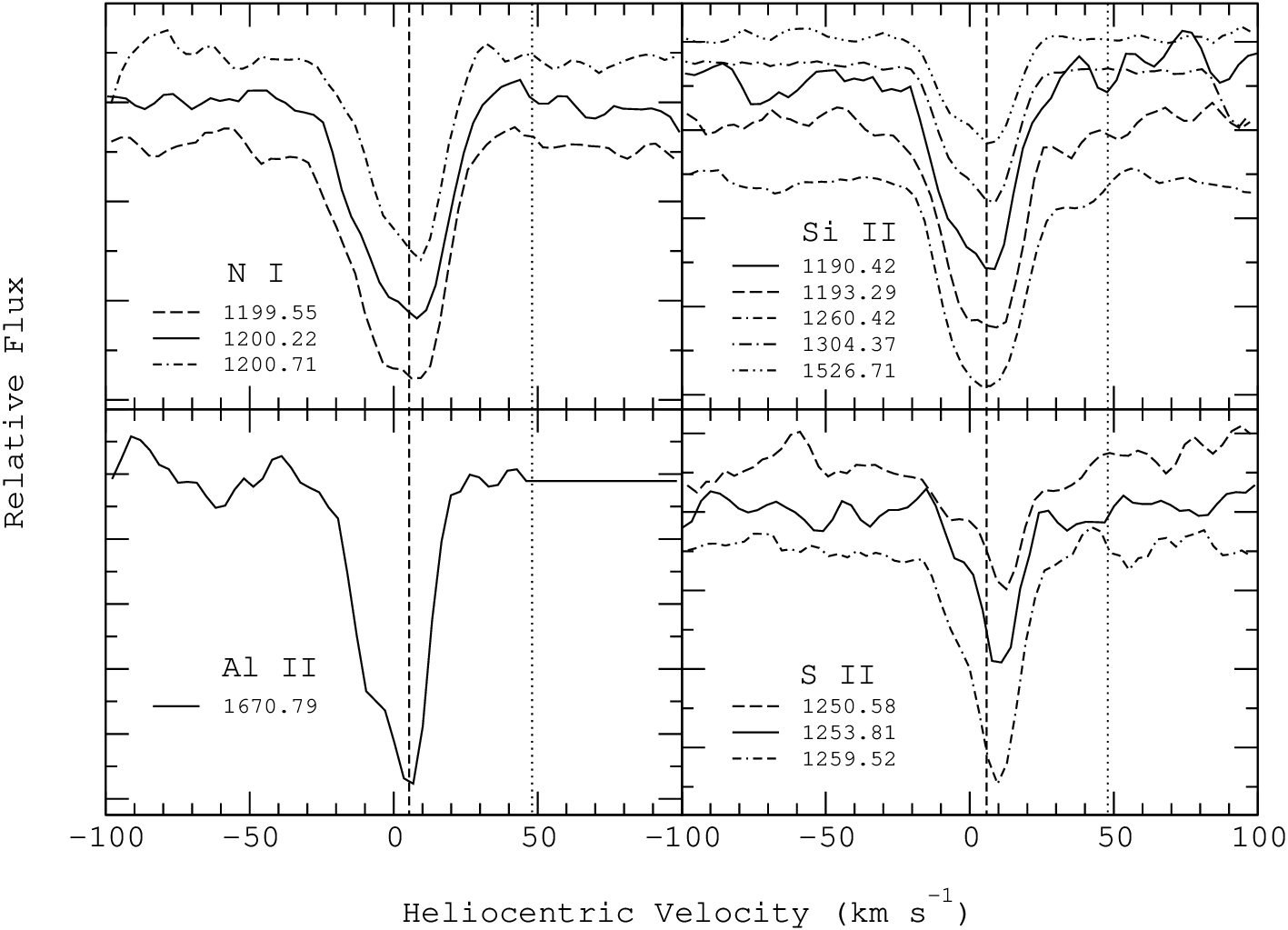}
\caption{Line velocity profiles for the \ion{N}{1} triplet, lines
from several \ion{Si}{2} multiplets, the \ion{Al}{2}$\lambda1670$
line, and the \ion{S}{2} triplet found in the IX~Vel spectrum
($S/N\sim30$).  The dashed line at $5.6\kms$ represents the LIC
velocity, while the dotted line at $48\kms$ represents the systemic
velocity.  The velocity centers of the absorption lines clearly match
the calculated LIC velocity.  \label{ix_vel}}
\end{figure}

\begin{figure}
\plotone{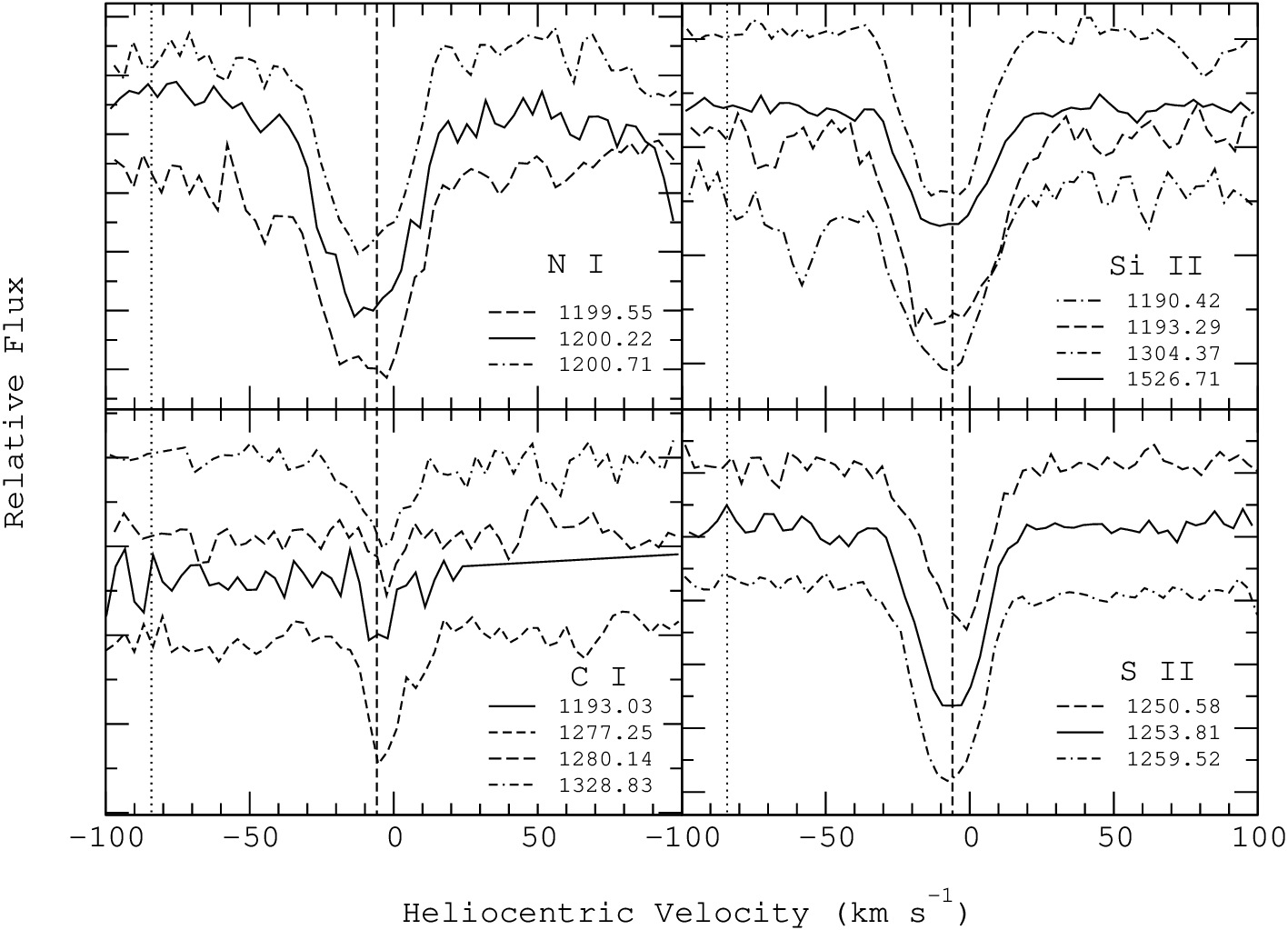}
\caption{Line velocity profiles for the \ion{N}{1} triplet, lines
from several \ion{Si}{2} and \ion{C}{1} multiplets, and the \ion{S}{2}
triplet appearing in the QU~Car spectrum ($S/N\sim20$).  The dashed
line at $-6.1\kms$ represents the LIC velocity, while the dotted line
at $-84\kms$ represents the systemic velocity.  The velocity line
centers are in good agreement with the calculated LIC
velocity.\label{qu_vel}}
\end{figure}

\begin{figure}
\plotone{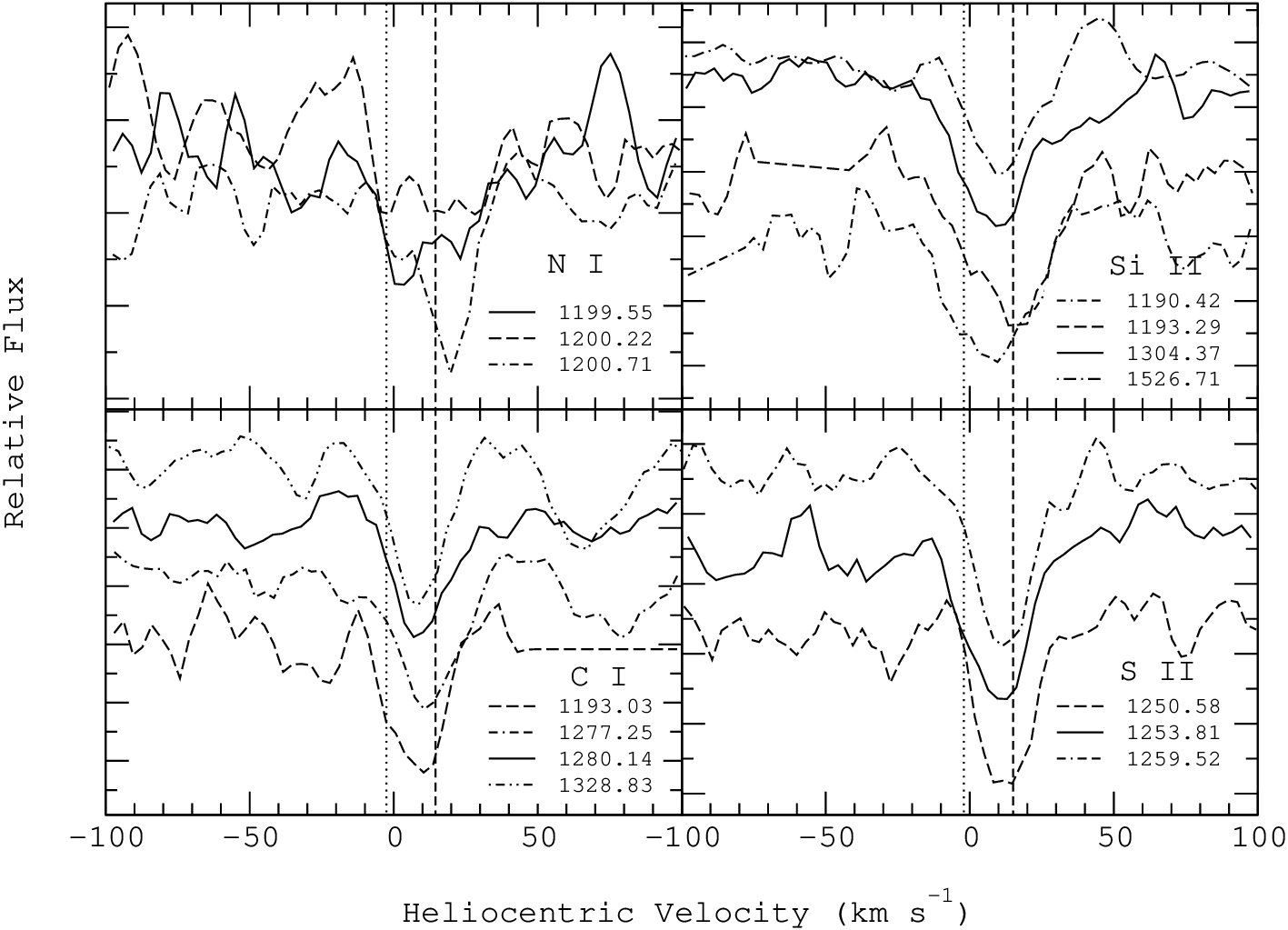}
\caption{Line velocity profiles for the \ion{N}{1} triplet, lines
from several \ion{Si}{2} and \ion{C}{1} multiplets, and the \ion{S}{2}
triplet appearing in the TZ~Per spectrum ($S/N\sim10$).  The dashed
line at $14.7\kms$ represents the LIC velocity, while the dotted line
at $-2\kms$ represents the systemic velocity.  Both the systemic and
LIC velocities fall within the velocity widths of the absorption
lines, but it is the LIC velocity that most closely matches the
line centers. \label{tz_vel}}
\end{figure}

\begin{figure}
\plotone{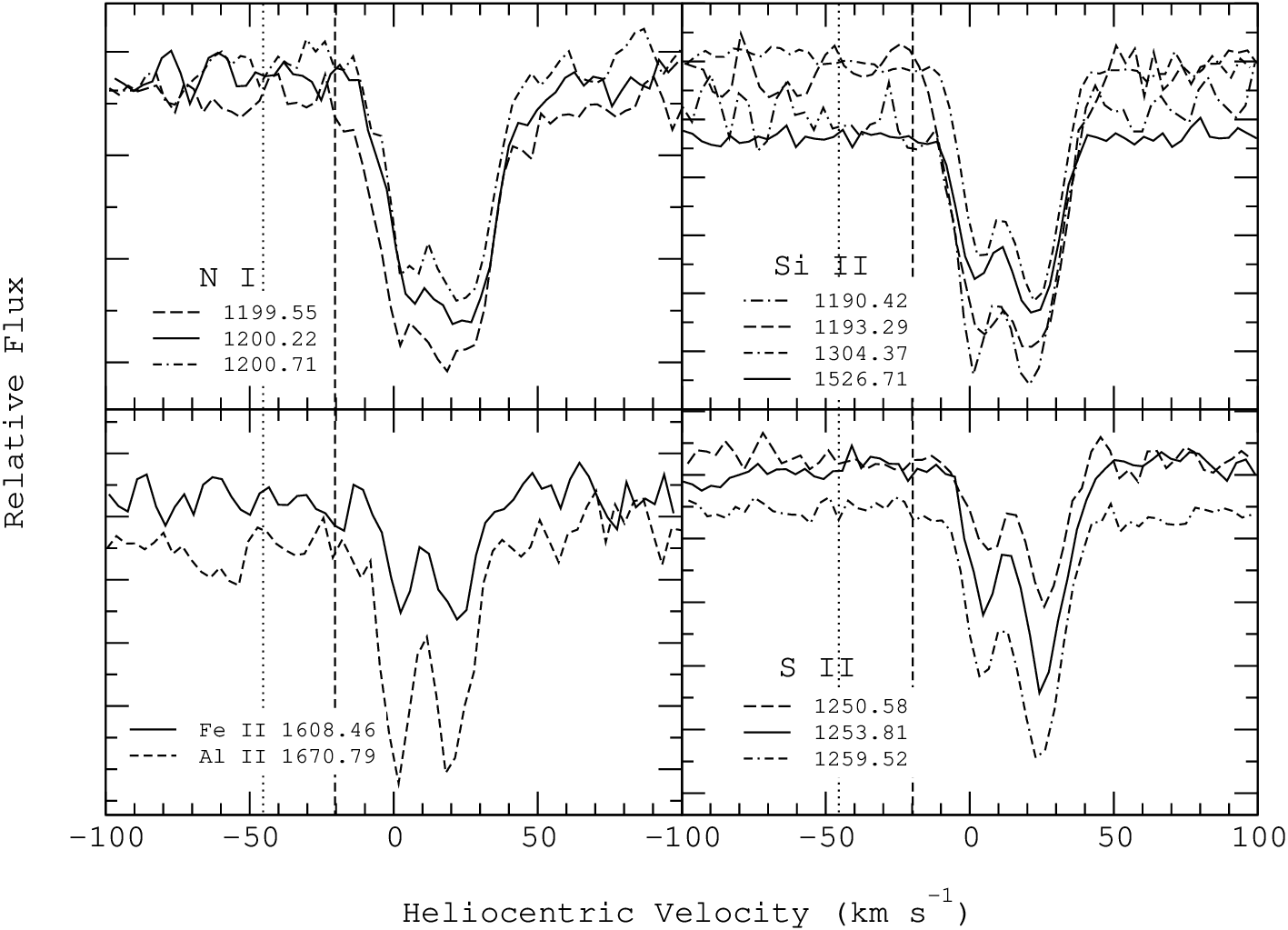}
\caption{Line velocity profiles for the \ion{N}{1} triplet, lines
from several \ion{Si}{2} multiplets, the \ion{Fe}{2} $\lambda1608$ and
\ion{Al}{2}$\lambda1670$ lines, and the \ion{S}{2} triplet appearing
in the V3885~Sgr spectrum ($S/N\sim30$).  The dashed line at
$-20.0\kms$ represents the LIC velocity, while the dashed line at
$-45\kms$ represents the systemic velocity.  Each absorption line
exhibits a two velocity component structure.
\label{v3_vel}}
\end{figure}

\begin{figure}
\plotone{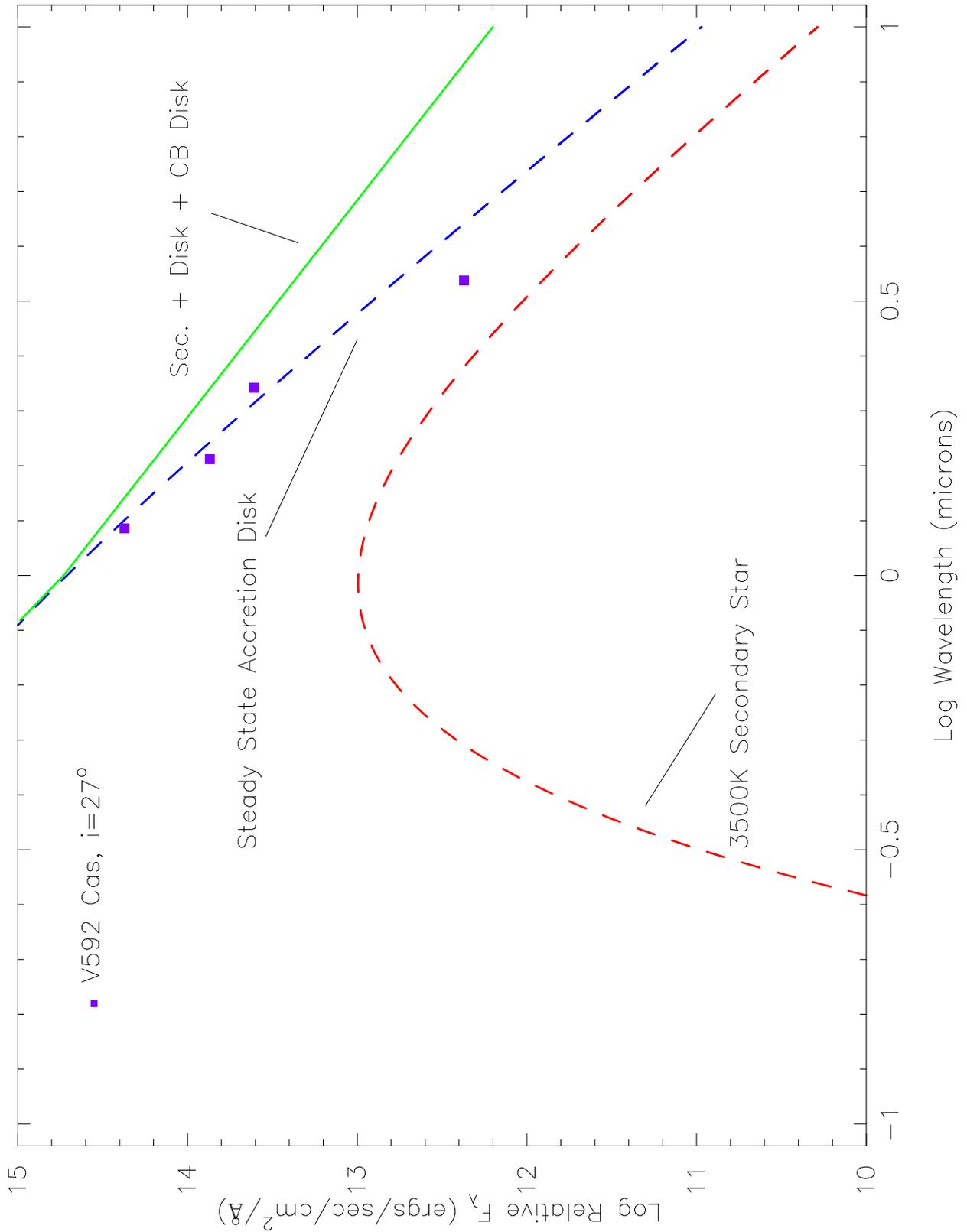}
\caption{A blackbody model of the IR flux of V592 Cas, including
the contribution from a CB disk, created using a steady state CV
accretion disk code.  The model for the CB disk is a ``sum of
blackbodies'' and the flux peaks near $3\,\micron$.  Also plotted are
the $JHKL'$ data points for V592~Cas, which do not match the CB disk
blackbody curve. \label{fred}}
\end{figure}

We have indicated the observational upper limit of $\nh \sim
10^{17}\cm^{-2}$ as the dashed horizontal line in Figure
\ref{diskmodel} for comparison of the observational numerical values
with the theoretical disk models.  The vertical dotted line at
$z/H\sim7$ represents the value of $z/H$ for EX Hya, while the
vertical dotted line at $z/H\sim17$ represents the value of $z/H$ for
IX~Vel, QU~Car, and V3885~Sgr (the values for DI~Lac and V592~Cas are
past the limits of the plot).  If CB disks exist, we can eliminate
Models C and D as the proposed CB disk structure for the four CVs,
as the model column densities are significantly above the detection
limit and absorption by such disk structures would be detectable at
the given inclinations.  A CB disk with the structure of Model A would
be detectable for the inclination of EX Hya and can therefore be ruled
out, as no absorption from a CB disk is seen in EX Hya.  Model B would
not be observed in any of our CV systems, but we note that this model
is the most conservative disk assumption of the four and we do not
expect it to be physically realistic (i.e., we would expect the CB
disk to have some form of an atmosphere).

As this investigation has only considered a small sample of CV systems
that could contain a CB disk, we plan to continue our photometric and
spectroscopic investigation for CB disks.  In particular, we plan to
analyze additional archival STIS data and to continue to look for 
their signatures in the IR SEDs of candidate CVs.

\acknowledgments 
We would like to thank D. Ciardi for the use of his $L'$ band image of
V592~Cas prior to publication.  KEB would like to thank R. Taam for
helpful discussions regarding CB disks.  We also thank the anonymous
referee for comments that greatly helped to improve this manuscript.
NS acknowledges support from NSF/REU grant AST-9810770.  SBH
acknowledges support from NSF grant AST-9810770.

\newpage

\clearpage
\newpage

\begin{deluxetable}{lllllll}
\tabletypesize{\scriptsize}
\tablewidth{0pt}
\tablecolumns{7}
\tablecaption{Object Log \label{objects}}
\tablehead{
\colhead{} & \colhead{$l$, $b$} & \colhead{Inclination ($z/H$)} & 
\colhead{LIC Velocity} & 
\colhead{Distance} & \colhead{Systemic Velocity} &  \colhead{} \\
\colhead{Object (type\tablenotemark{a} )} & \colhead{(deg)} & \colhead{(deg)}& 
\colhead{($\kms$)} & \colhead{(pc)} & \colhead{($\kms$)} &  
\colhead{References} }
\startdata 
DI~Lac (N)    & $103.107$, $-4.855$ & $18$ (42) & \phs\phn$3.5$     & $2500$
              & \phs$1$   & 1,2 \\
EX~Hya (IP)   & $303.186$, $33.622$ & $78\pm1$ (7) & $-14.8$  & $64.5\pm1.2$ 
              & $-180$ & 3,4\\
IX~Vel (NL)   & $264.929$, $-7.890$ & $60\pm5$ (17.6) &\phs\phn$5.6$&$95\pm12$ 
              & \phs$48$  & 5,6\\
QU~Car (NL)   & $293.513$, $-7.717$ & $60$ (17.6) & \phn$-6.1$   & $2000$   
              & $-84\pm20$ & 7,8 \\
TZ~Per (DN)   & $133.576$, $-2.777$ & \nodata & \phs$14.7$     & $80$  
              & $-2\pm2$  & 9,10\\
V3885~Sgr (DN)&$357.477$, $-27.762$ & $50-70$ (17.6) & $-20.0$ & $110\pm25$  
              & $-45\pm4$ & 5,11,12\\
V592 Cas (NL) & $118.603$, $-6.910$ & $28\pm11$ (36) & \phs\phn$9.9$ &  $60$
              & \phs$21\pm14$    & 13
\enddata
\tablerefs{(1)~\citet{moy03}; (2)~\citet{kra64}; (3)~\citet{hel87};
(4)~\citet{beu03};
(5)~\citet{beu90}; (6)~\citet{per97}; (7)~\citet{gill82}; (8)~\citet{dre03}; 
(9)~\citet{ech99}; (10)~\citet{ber85}; (11)~\citet{cow77}; (12)~\citet{hau85};
(13)~\citet{hub98} }
\tablenotetext{a}{N=nova, IP=intermediate polar, NL=nova-like, DN=dwarf novae}
\end{deluxetable}

\begin{deluxetable}{cccccc}
\tablewidth{0pt}
\tablecolumns{5}
\tablecaption{HST Observations Log \label{log}}
\tablehead{
\colhead{} & \colhead{} & \colhead{} & 
\colhead{} & \colhead{Exposure Time} & \colhead{} \\
\colhead{Object} & \colhead{Data Set}  & 
\colhead{Observation Date} & \colhead{Grating} & \colhead{(s)} & 
\colhead{S/N}}
\startdata 
DI~Lac      &  O5B607010 & 2000 April 19 & E140M & 2127 & 5\\
EX~Hya      &  O68301010\tablenotemark{a} & 2000 May
		& E140M & 15,200 & 20  \\
IX~Vel      &  O5BI01010 & 2000 April 3 & E140M & 1750 & 30\\
QU~Car      &  O5BI08010 & 2000 July 21 & E140M & 2600 & 20 \\
TZ~Per      &  O5B610010 & 2000 July 3 & E140M & 2151 & 10 \\
V3885~Sgr   &  O5BI04010 & 2000 April 30 & E140M & 2480 & 30\\
\enddata
\tablenotetext{a}{The EX~Hya data sets O68301010, O68301020, O68301030, 
O68302010, O68302020, and O68302030 were used for the EX~Hya measurements.} 
\end{deluxetable}

\begin{deluxetable}{llcccccccccccc} 
\tabletypesize{\footnotesize}
\rotate
\tablewidth{0pt}
\tablecolumns{14}
\tablecaption{Identified Absorption Lines \label{all_tab} }
\tablehead{
\colhead{} & \colhead{} & \multicolumn{2}{c}{DI~Lac} & 
\multicolumn{2}{c}{EX~Hya}
& \multicolumn{2}{c}{IX~Vel} & \multicolumn{2}{c}{QU~Car} & 
\multicolumn{2}{c}{TZ~Per} & \multicolumn{2}{c}{V3885~Sgr} \\
\multicolumn{2}{c}{Ion $\lambda_{\rm rest}$} & \colhead{RV} & 
\colhead{EW\tablenotemark{a}} & 
\colhead{RV}& \colhead{EW} & \colhead{RV}& \colhead{EW} & \colhead{RV} &
\colhead{EW} & \colhead{RV}& \colhead{EW} & \colhead{RV} & \colhead{EW}  \\
\multicolumn{2}{c}{(\AA)} & \colhead{($\kms$)}& \colhead{(m\AA)} & 
\colhead{($\kms$)}&
\colhead{(m\AA)} & \colhead{($\kms$)} & \colhead{(m\AA)}
& \colhead{($\kms$)}& \colhead{(m\AA)} & \colhead{($\kms$)}& \colhead{(m\AA)} 
& \colhead{($\kms$)}& \colhead{(m\AA)} }
\startdata


\ion{N}{1} & $\lambda1159.82$
	& \nodata&\nodata & \nodata&\nodata & \nodata&\nodata & $-12.28$
	& $43$ &\nodata & \nodata&\nodata &\nodata \\
\ion{C}{1} & $\lambda1188.83$
	&\nodata &\nodata &\nodata &\nodata &\nodata &\nodata & $-8.00$
	& $12$ &\nodata &\nodata &\nodata &\nodata \\
\ion{C}{1} & $\lambda1190.02$
	&\nodata &\nodata &\nodata &\nodata &\nodata &\nodata &\nodata
	&\nodata & $13.80$ & $45$ &\nodata &\nodata \\
\ion{Si}{2} & $\lambda1190.42$ 
	&\nodata&\nodata & $-10.08$&$122$ & $3.81$ & $93$ & $-10.53$
	&$147$ & $6.24$ & $125$ & $1.00,23.68$ & $62,84$ \\
\ion{Si}{2} & $\lambda1193.29$ 
	& \nodata&\nodata& $-8.80$&$103$ & $3.78$ & $112$ & $-8.02$ 
	& $173$ & $15.06$ & $84$ & $6.09,21.17$ & $98,116$  \\
\ion{C}{1} & $\lambda1193.03$
	&\nodata &\nodata &\nodata &\nodata &\nodata &\nodata & $-2.99$
	& $7$ & $13.81$ & $91$ &\nodata &\nodata \\
\ion{Mn}{2} & $\lambda1197.18$
	&\nodata &\nodata &\nodata &\nodata &\nodata &\nodata &$-5.53$
	& $27$ &\nodata &\nodata &\nodata &\nodata \\
\ion{Mn}{2}    & $\lambda1199.39$
	&\nodata &\nodata &\nodata &\nodata &\nodata &\nodata &$-3.04$
	& $16$ &\nodata &\nodata &\nodata &\nodata \\ 	
\ion{Mn}{2}    & $\lambda1201.12$
	&\nodata &\nodata &\nodata &\nodata &\nodata &\nodata &$-8.05$
	&$16$ &\nodata &\nodata &\nodata &\nodata \\
\ion{N}{1}  & $\lambda1199.55$ 
	&\nodata&\nodata & $-5.60$&$115$ & $1.31$ & $138$& $-10.45$
	& $182$ & $13.82$ & $110$ & $3.69,21.20$ & $93,117$  \\
\ion{N}{1}    & $\lambda1200.22$
	& \nodata&\nodata & $-5.90$ &$105$ & $2.88$ & $126$ & $-11.38$  
	&$156$ & $16.32$ & $89$ & $6.20,23.70$ & $74,95$ \\
\ion{N}{1}    & $\lambda1200.71$ 
	&\nodata&\nodata & $-7.67$&$78$ & $3.75$&$100$ & $-10.55$ 
	&$131$ & $18.44$&$127$ & $6.21,23.70$ & $74,105$   \\
\ion{Si}{3} & $\lambda1206.50$ 
	& \nodata&\nodata & $-3.13$ &$68$ & $11.10$ &$73$ & $-3.10$ 
	&$116$ & \nodata&\nodata & $11.28$ & $76$ \\
\ion{Mg}{2} & $\lambda1239.93$
	&\nodata &\nodata &\nodata &\nodata &\nodata &\nodata &$-5.79$	
	&$26$ & $11.48$ &$36$ &\nodata &\nodata \\
\ion{Mg}{2} & $\lambda1240.39$
        & \nodata&\nodata & \nodata&\nodata&\nodata&\nodata& $-0.96$ 
	& $19$ & \nodata&\nodata & \nodata&\nodata \\
\ion{S}{2}  & $\lambda1250.58$ 
	& $-1.64$&$125$ & $-9.01$&$11$ & $7.04$&$43$ & $-6.81$
	&$91$ & $8.16$ & $94$ & $5.81,26.20$& $21,24$ \\
\ion{S}{2}    & $\lambda1253.81$ 
	& $-9.03$&$233$ &$-8.03$&$35$ & $10.11$&$27$ & $-6.11$
       	&$105$ & $6.51$&$99$ & $4.67,26.20$& $24,48$\\
\ion{S}{2}    & $\lambda1259.52$ 
	& $-3.49$&$182$ & $-7.98$&$38$ & $8.12$&$44$ & $-5.67$
	&$111$ & $11.79$&$109$ & $2.38,23.82$ & $26,56$  \\
\ion{Si}{2} & $\lambda1260.42$ 
	& \nodata&\nodata & $-8.86$&$8$ & $4.99$ & $136$ & $0.73$
	&$269$ &$18.84$ & $228$ & $7.16,23.82$& $115,126$  \\
\ion{C}{1} & $\lambda1260.74$
	&\nodata &\nodata &\nodata &\nodata &\nodata &\nodata &\nodata
	&\nodata &$9.18$ &$60$ &\nodata &\nodata \\
\ion{C}{1} & $\lambda1276.48$
        & \nodata&\nodata & \nodata&\nodata&\nodata&\nodata&\nodata
        & \nodata& $-3.19$ & $32$&\nodata&\nodata \\
\ion{C}{1} &$\lambda1277.25$
        & $-1.49$ &$96$ &\nodata &\nodata &\nodata &\nodata &$-1.31$
	& $33$ & $13.97$ & $21$ &\nodata &\nodata \\ 
\ion{C}{1}$^*$    & $\lambda1277.55$
        & \nodata&\nodata & \nodata&\nodata&\nodata&\nodata&\nodata
        & \nodata& $6.93$ & $50$ &\nodata&\nodata \\
\ion{C}{1} & $\lambda1280.14$
	& $1.47$& $56$ &\nodata &\nodata & \nodata &\nodata &$1.33$
	&$6$ & $13.98$ & $81$ &\nodata &\nodata \\
\ion{P}{2} & $\lambda1301.87$
	& $7.86$&$64$ & \nodata&\nodata & \nodata&\nodata& $-3.83$
	& $20$ & $9.41$&$46$ &\nodata&\nodata \\
\ion{O}{1} & $\lambda1302.17$ 
	& $-12.87$&$196$ & $-9.31$&$136$ & $0.43$&$163$ &$-8.44$
	&$195$ & $2.96$&$289$ & $12.38,51.54$ & $195,17$  \\
\ion{Si}{2} &$\lambda1304.37$
        & $-3.64$ &$275$ &$-8.28$ & $79$ & $5.01$ &$81$ &$-8.45$
	&$135$ & $11.72$ &$140$ & $3.20,23.90$ &$52,86$ \\
\ion{Ni}{2} & $\lambda1317.22$ 
	& $-5.83$&$74$ & \nodata&\nodata & \nodata&\nodata & $-3.26$
	& $20$ & \nodata&\nodata & \nodata&\nodata  \\
\ion{C}{1}  & $\lambda1328.83$ 
	& $-1.21$ & $91$ & \nodata&\nodata & \nodata&\nodata & $-4.02$
	&$22$ & $9.55$ &\nodata & \nodata&\nodata  \\
\ion{C}{1}$^*$    & $\lambda1329.09$
	& $3.30$&$75$ & \nodata&\nodata&\nodata&\nodata&\nodata 
	&\nodata& $11.81$ & $65$ & \nodata&\nodata \\
\ion{C}{2}  & $\lambda1334.53$ 
	& $-6.20$ & $44$ & $-10.79$&$151$ & $1.93$&$170$ & $-6.82$
	& $213$ & $11.82$&$244$ & $12.20$&$214$  \\
\ion{C}{2}$^*$    & $\lambda1335.66$ 
	& $4.95$&$166$ & $0.36$&$28$ & $5.20$&$54$ & $-10.36$
	&$129$ & $7.79$&$118$ & $14.97,32.94$&$25,33$  \\
\ion{Cl}{1}  & $\lambda1347.24$
	& \nodata&\nodata & \nodata&\nodata&\nodata&\nodata& $-6.28$
	& $27$ & $7.41$ & $92$&\nodata&\nodata \\
\ion{Ni}{2} & $\lambda1370.13$ 
	& $5.56$&$177$ & \nodata&\nodata & \nodata&\nodata & \nodata
	&\nodata & \nodata&\nodata & \nodata&\nodata  \\
\ion{S}{1} & $\lambda1425.03$
	&\nodata &\nodata &\nodata &\nodata &\nodata &\nodata &\nodata
	&\nodata & $3.69$ &$21$ &\nodata &\nodata \\
\ion{S}{1}    & $\lambda1425.19$
	& \nodata&\nodata & \nodata&\nodata &\nodata& \nodata&\nodata
	& \nodata & $5.80$ &$18$&\nodata&\nodata \\
\ion{Si}{2} & $\lambda1526.71$ 
	& $-3.60$ &$525$ &$-8.21$ &$129$ & $3.67$ &$111$ & $-6.49$ 
	&$172$ & $5.12$&$153$ & $0.65,24.23$ &$73,106$  \\
\ion{P}{2} & $\lambda1532.53$ 
	& \nodata&\nodata & \nodata&\nodata &\nodata&\nodata&\nodata
	& \nodata& $10.45$ & $82$&\nodata&\nodata \\
\ion{C}{1}$^*$ & $\lambda1560.70$
	& $-2.13$&$110$ &\nodata&\nodata&\nodata&\nodata&\nodata
	&\nodata&\nodata&\nodata&\nodata&\nodata \\
\ion{Fe}{2} & $\lambda1608.46$ 
	& $5.63$ & $260$ & $-7.46$ & $42$ & $2.40$ & $49$ & $-7.45$
	& $131$& $6.99$&$125$&$1.95,22.47$ & $23,24$ \\
\ion{Al}{2} & $\lambda1670.79$
	& \nodata&\nodata & $-8.02$&$138$ & $-11.25$&$61$ &$1.80$
	&$161$ & $16.79$ & $145$ & $1.06,20.81$ & $45,51$ \\
\enddata
\tablenotetext{a}{Lines with EW $\,>100$ m\AA\ are likely saturated or
blended components; we therefore assign an error of $\pm7\kms$ to the
radial velocities of these lines.  See text for details.}

\end{deluxetable}


\newpage
\begin{thebibliography}{99}
%
\bibitem[Bell et al.(1997)]{bel97}
Bell, K. R., Cassen, P. M., Klahr, H. H., \& Henning, T.
1997, \apj, 486, 372
%
\bibitem[Berriman, Szkody, \& Capps(1985)]{ber85}
Berriman, G., Szkody, P., \& Capps, R. W.
1985, \mnras, 217, 327
%
\bibitem[Beuermann et al.(2003)]{beu03}
Beuermann, K. Harrison, T. E., McArthur, B. E., Benedict, G. F., \&
G\"{a}nsicke, B. T.
2003, \aap, 412, 821
%
\bibitem[Beuermann \& Thomas(1990)]{beu90}
Beuermann, K. \& Thomas, H.-C. 
1990, \aap, 230, 326
%
\bibitem[Cassen(1993)]{cas93}
Cassen, P. M. 
1993, Lunar \& Planet. Sci. Conf., XXIV, 261
%
\bibitem[Ciardi et al.(1998)]{cia98}
Ciardi, D. R., Howell, S. B., Hauschildt, P. H., \& Allard. F.
1998, \apj, 504, 450
%
\bibitem[Cowley et al.(1977)]{cow77}
Cowley, A. P., Crampton, D., \& Hesser, J. E. 
1977, \apj, 214, 471
%
\bibitem[Drew et al.(2003)]{dre03}
Drew, J. E., Hartley, L. E., Long, K. S., \& van der Walt, J.
2003, \mnras, 338, 401
%
\bibitem[Dubus et al.(2002)]{dub02}
Dubus, G., Taam, R. E., \& Spruit, H. C. 
2002, \apj, 569, 395
%
\bibitem[Echevarr\'{i}a et al.(1999)]{ech99}
Echevarria, J., Pineda, L., \& Costero, R.
1999, RMxAA, 35, 135
%
\bibitem[Frank, King, \& Raine(2002)]{fkr02}
Frank, J., King, A. R., \& Raine, D. J.
2002, Accretion Power in Astrophysics (3d ed.; Cambridge, Cambridge
University Press)
%
\bibitem[Frisch, Grodnicki, \& Welty(2002)]{fri02}
Frisch, P. C., Grodnicki, L., \& Welty, D. E.
2002, \apj, 574, 834 (FGW)
%
\bibitem[Gilliland \& Phillips(1982)]{gill82}
Gilliland, R. L. \& Phillips, M. M. 
1982, \apj, 261, 617 
%
\bibitem[Haug \& Drechsel(1985)]{hau85}
Haug, K. \& Drechsel, H. 
1985, \aap, 151, 157
%
\bibitem[Hellier et al.(1987)]{hel87}
Hellier, C. et al. 
1987, \mnras, 228, 463
%
\bibitem[Holberg, Barstow, \& Sion(1998)]{hol98}
Holberg, J. B., Barstow, M. A., \& Sion, E. M.
1998, \apjs, 119, 207
%
\bibitem[Howell et al.(2001)]{how01}
Howell,  S. B., Nelson, L., \& Rappaport, S. 
2001, \apj, 550, 518
%
\bibitem[Huber et al.(1998)]{hub98}
Huber, M. E., Howell, S. B., Ciardi, D. R., \& Fried, R.
1998, \pasp, 110, 784
%
\bibitem[Kimura, Mann, \& Jessberger(2003)]{kim03}
Kimura, H., Mann, I., Jessberger, E. K.
2003, \apj, 582, 846
%
\bibitem[Kraft(1964)]{kra64}
Kraft, R. P. 
1964, \apj, 139, 457
%
\bibitem[Lallement \& Bertin(1992)]{lal92}
Lallement, R. \& Berlin, P.
1992, \aap, 266, 479
%
\bibitem[Morton(1991)]{mor91}
Morton, D. C.
1991, \apjs, 77, 119
%
\bibitem[Moyer et al.(2003)]{moy03}
Moyer, E., Sion, E. M., Szkody, P., G\"{a}nsicke, B., Howell, S., \&
Starrfield, S.
2003, \apj, 125, 288
%
\bibitem[Perryman et al.(1997)]{per97}
Perryman, M. A. C., et al.
1997, \aap, 323, 49
%
\bibitem[Spruit \& Taam(2001)]{spr01}
Spruit, H. C. \& Taam, R. E.
2001, \apj, 548, 900
%
\bibitem[Taam, Sandquist, \& Dubus(2003)]{taa03}
Taam, R. E., Sandquist, E. L., Dubus, G.
2003, \apj, 592, 1124
%
\bibitem[Taam \& Spruit(2001)]{taa01}
Taam, R. E. \& Spruit, H. C. 
2001, \apj, 561, 329
%
\bibitem[Wood et al.(2002)]{woo02}
Wood, B. E., Redfield, S., Linsky, J. L., \& Sahu, M. S.
2002, \apj, 581, 1168
%
\end{thebibliography}
\end{document}